# Casimir force among spheres made of Weyl semimetals breaking Lorentz reciprocity

Yoichiro Tsurimaki,[*] Xin Qian, Simo Pajovic, Svetlana V. Boriskina,[*] and Gang Chen[*]

Department of Mechanical Engineering, Massachusetts Institute of Technology, Cambridge, MA, 02139

**Abstract**

The Casimir force and thermal Casimir force originating from quantum electromagnetic fluctuations at zero and non-zero temperatures, respectively, are significant in nano- and microscale systems and are well-understood. Less understood, however, are the Casimir and thermal Casimir forces in systems breaking Lorentz reciprocity. In this work, we derive a formalism for thermal Casimir forces between an arbitrary number of spheres based on fluctuational electrodynamics and scattering theory without the assumption of Lorentz reciprocity. We study the total Casimir force in systems of two and three Weyl semimetal spheres with time-reversal symmetry breaking for different orientations of the momentum-space separation of Weyl nodes in both thermal equilibrium and nonequilibrium. In thermal nonequilibrium, we show that a net thermal Casimir force exists not only along the center-to-center displacements of the spheres, but also in the transverse direction to it due to thermal emission with non-zero angular momentum. Different symmetries of the system drive a variety of dynamics such as global rotations, self-propulsion, and spinning of the spheres. We also show that the Casimir energy in thermal equilibrium depends on the orientations of the Weyl node directions in the spheres and that the lateral Casimir force will act between the spheres even in thermal equilibrium to relax the system into the minimum energy state without transferring net energy and momentum to the environment. The developed framework opens a way for investigating many-body dynamics by Casimir and thermal Casimir forces among arbitrary number of spheres with arbitrary dielectric function tensors in both thermal equilibrium and nonequilibrium.

---

[*] E-mail correspondence: ytsuri@mit.edu, gchen2@mit.edu, sborisk@mit.edu

# I. Introduction

Quantum (zero-point energy) and thermal fluctuations of electromagnetic fields in materials lead to force and radiative heat transfer. At micro- and nanoscales, the characteristic distance between objects that exchange energy radiatively becomes shorter than the mean wavelength of thermal radiation near room temperature. At such near-field, radiative heat transfer between objects in thermal nonequilibrium exceeds Planck's law of thermal radiation in the far-field [1-7] due to tunneling of photons, and phonons [8]. Similarly, the van der Waals force and Casimir force [9-13], as well as Casimir torque [14-17], due to zero-point fluctuations of electromagnetic fields, become significant at these scales, reflecting the increased near-field transfer of both linear and angular momenta of light. Thermal Casimir forces due to thermal fluctuations of electromagnetic fields, are also important at this scale [18], especially at longer length scales. The focus of most work on the fluctuation-induced radiative heat and momentum transfer, however, has been on systems that obey the Lorentz reciprocity theorem of electromagnetism, i.e., reciprocal systems.

Recently, the scope of near-field radiative heat transfer in many-body systems was extended to systems breaking Lorentz reciprocity, i.e., nonreciprocal systems, by breaking time-reversal symmetry [19,20]. Most well-studied systems with broken time-reversal symmetry comprise gyroelectric materials: magnetic semiconductors such as InSb [19-22] under an external magnetic field or magnetic materials with large anomalous Hall angles such as magnetic Weyl semimetals [23,24]. While the optical properties of prototypical magnetic Weyl semimetals were used in the previous theoretical proposals, several realistic magnetic Weyl semimetals such as $Co_2MnX$ (X=Ga, Al) [25,26] are promising for above room temperature applications. In such gyroelectric materials, the degeneracy of certain influential optical modes with non-zero angular momentum, e.g., surface plasmon polaritons, is lifted and the frequency dispersion of the modes becomes nonreciprocal. Different photon occupation numbers on these nonreciprocal modes result in thermal radiation from such surfaces carrying non-zero angular momentum [27,28]. As a consequence, thermal radiation from a nonreciprocal object exhibits a strong angular momentum flux in the near-field.

In many-body systems where the angular-momentum-carrying thermal radiation from nonreciprocal objects interact in the near-field, interesting yet counterintuitive phenomena such as persistent heat current in thermal equilibrium [19,22,29], photon thermal Hall effect [22,24,30], anisotropic thermal magnetoresistance [31], thermal rectification [21], and thermal routing [23] have been predicted. Moreover, by coupling these nonreciprocal modes in the near-field to the far-field, the classical Kirchhoff's law of radiation, or the equality of spectral directional absorptivity and emissivity, can be violated [32-36]. These nonreciprocal phenomena provide new opportunities for controlling radiative heat transfer, which promises new thermal devices such as rectifiers, routers, and magnetic sensors based on thermal radiation.

The momentum carried by radiation in nonreciprocal systems can also be utilized for manipulating micro- or nanoscale objects beyond what is allowed by reciprocal systems. Typically, Casimir forces between two identical objects in thermal equilibrium are attractive. Casimir forces in nonreciprocal systems can go beyond the typical attractive force. For example, repulsive Casimir forces were predicted to exist between magnetic Weyl semimetal slabs [37,38], two reciprocal materials separated by a chiral medium [39], and photonic topological insulators [40] by evading the no-go theorem [41] that restricts the force to be attractive in mirror-symmetric systems. The lateral Casimir force that is transverse to the separation direction was also predicted on a two-level system when it is situated above a nonreciprocal medium [42-44] that hosts nonreciprocal surface modes.

Thermal Casimir forces on objects and the environment originate from their thermal radiation, which depend on their temperatures and material properties. In thermal nonequilibrium, characteristics of thermal radiation from the objects and the environment can appear more clearly in thermal Casimir forces compared to thermal equilibrium. For instance, thermal radiation from a gyroelectric material carries angular momentum in the direction of the material's gyration vector and exhibits circulating Poynting vector distribution around the object [27]. Thus, an object in proximity to the gyroelectric material will experience lateral thermal Casimir force. The understanding of the dependence of Casimir and thermal Casimir forces on the temperatures, material properties, and orientations of the nonreciprocal objects will be crucial for active mechanical control of the objects in ways that are not possible in reciprocal systems. However, thermal Casimir forces remain largely unexplored for systems involving nonreciprocal objects. Only recently, a lateral thermal Casimir force was shown to arise between reciprocal and nonreciprocal objects out of thermal equilibrium, which can output work as a heat engine [45].

The transfer of Casimir torque in nonreciprocal systems was also studied for different geometries including two macroscopic slabs [46], a two-level system above nonreciprocal slab [47], and the transfer of thermal Casimir torque was studied in thermal nonequilibrium between a single nanosphere and the environment [48,49], and between two nanocubes and the environment [50]. It was shown that the net transfer of thermal Casimir torque between the objects and the environment in thermal nonequilibrium results in non-zero torques on the objects due to the conservation of angular momentum.

Furthermore, Casimir forces between two objects can exhibit nonmonotonic dependence on the separation when a third object is in their proximity and the pairwise approximation can break down [51-53]. Effects of nonreciprocity on such many-body Casimir interactions are of interest from a fundamental physics point of view. Additionally, the dynamics of a collection of nanoscale objects through Casimir and thermal Casimir forces are important for engineering systems such as assembly of nanoparticles, and nonreciprocity can introduce new degrees of freedom. However, Casimir interactions in many-body systems comprising nonreciprocal objects are not explored before. Although creating different gyration directions in magneto-optical objects by external magnetic fields may be challenging at nano- and microscales, magnetic materials with exceptionally large anomalous Hall angle, e.g., magnetic Weyl semimetals [54], may provide a promising platform for considering many-body objects with different gyration directions by changing the crystal orientations of each object in the system.

To study Casimir interactions in many-body nonreciprocal systems, however, an exact and computationally efficient formalism of Casimir and thermal Casimir forces is necessary. Previously, the scattering formalism [55] was developed to calculate the total Casimir forces, i.e., both Casimir and thermal Casimir forces, in thermal equilibrium at finite temperatures for arbitrary number of materials with arbitrary material properties. However, such formalism of Casimir forces in thermal nonequilibrium is not available. We address this challenge by deriving compact formulas for the thermal Casimir forces for systems consisting of an arbitrary number of spheres with arbitrary dielectric permittivity tensors in thermal nonequilibrium. The obtained formula is exact within the framework of the scattering formalism and fluctuational electrodynamics. With the developed formalism, combined with the scattering formalism [55] for total Casimir forces in thermal equilibrium, we can calculate total Casimir forces between an arbitrary number of spheres with arbitrary dielectric function tensors in thermal equilibrium or nonequilibrium. We applied the developed method to study total Casimir forces in many-body systems composed of Weyl semimetal spheres with time-reversal symmetry breaking in thermal equilibrium and nonequilibrium for different object orientations relative to each other.

We show that a lateral thermal Casimir force exists between two magnetic Weyl semimetal spheres in thermal nonequilibrium for certain relative object orientations. While the force in the inter-sphere direction is contributed by both Casimir and thermal Casimir forces, the force in the lateral direction is solely due to the thermal Casimir force. This lateral thermal Casimir force was recently discussed between two bodies in vacuum and proposed as repulsion force for a heat engine [56]. We show that directions of the lateral thermal Casimir forces are restricted by symmetries of the gyration directions in the spheres and controllable by changing the orientations of the spheres. Different symmetries of the gyration axes of the spheres drive different dynamics such as global rotations, self-propulsion, and spinning of the spheres. We also study the total Casimir forces between a magnetic Weyl semimetal and a silica sphere to show that the lateral thermal Casimir force arises when at least one sphere is nonreciprocal. We also apply our formalism to a many-body system composed of three magnetic Weyl semimetal spheres with their gyration axes pointing normal to the inter-sphere direction. We show that the spheres experience a global rotation due to the lateral thermal Casimir force in thermal nonequilibrium situations. For both two-sphere and three-sphere systems, we identify the energetically favorable system configurations, and show that even in the case of *global thermal equilibrium*, a lateral Casimir force exists to relax the orientations of the spheres to the state with the minimum Casimir energy.

This paper is organized as follows. In Section II, we describe the nonreciprocal many-body system and review the scattering formalism of total Casimir forces in thermal equilibrium. Then, we describe the derivation of the formulas of the thermal Casimir force in thermal nonequilibrium. We introduce the dielectric function model of magnetic Weyl semimetals and silica used in this work. We also summarize the computational workflow. In section III, we discuss radiation fields around a magnetic Weyl semimetal sphere to highlight the non-zero angular-momentum of thermal emission in the near-field and to develop intuitions about thermal Casimir force interactions. We then discuss our results of total Casimir force and Casimir energy in the systems comprising of two magnetic Weyl semimetal spheres, a magnetic Weyl semimetal sphere and a silica sphere, and finally three magnetic Weyl semimetal spheres. In Section IV, we summarize the results and discuss potential applications of this work.

## II. Formalism of Casimir and thermal Casimir forces

### A. Description of the system

We first describe the system of interest and our underlying assumptions. We consider a system composed of $N$ nonintersecting spheres in vacuum as shown in Fig. 1 (a). We assume the electromagnetic responses of the spheres are linear, described by frequency-dependent, lossy dielectric function tensors $\bar{\bar{\varepsilon}}_j(\omega)$ ($j = 1, \cdots, N$). The relative magnetic permeability of all the spheres is assumed isotropic and taken to be unity $\mu_j = 1$ ($j = 1, \cdots, N$). We also assume the temperatures of the spheres are homogeneous and are denoted by $T_j$ ($j = 1, \cdots, N$), which can be different from each other and from the environment temperature $T_{env}$ in thermal nonequilibrium. In this work, we realize a nonreciprocal optical response by having gyroelectric materials in the system and we schematically represent a gyration axis of a sphere by an arrow overlapped on the sphere as shown in Fig. 1 (a). The dielectric function tensors of the gyroelectric spheres are not symmetric, $\bar{\bar{\varepsilon}}_j(\omega) \neq \bar{\bar{\varepsilon}}_j^T(\omega)$, and the system violates Lorentz reciprocity [57]. More generally, the formalisms developed in this work can be applied to an arbitrary form of dielectric function tensors. Particularly, we consider magnetic Weyl semimetals with a single pair of Weyl nodes as a model gyroelectric material [58]. In such magnetic Weyl semimetals, the direction of the momentum-space Weyl node separation corresponds to the gyration axis of the material in the real space as shown in Fig. 1 (b). The model of magnetic Weyl semimetals considered in this work will be discussed in section II. D.

We denote the total Casimir force acting on the $j$-th sphere in the system as $\boldsymbol{F}^{(j)}$. In general, the total Casimir force can be decomposed into the equilibrium contribution and the nonequilibrium contribution as [59]:

$$\boldsymbol{F}^{(j)}(T_{env}, T_1, \cdots, T_j, \cdots, T_N) = \boldsymbol{F}^{(j),eq}(T_{env}) + \sum_{k=1,\cdots,j,\cdots,N} \left[\boldsymbol{F}_k^{(j),th}(T_k) - \boldsymbol{F}_k^{(j),th}(T_{env})\right]$$
$$= \boldsymbol{F}^{(j),eq}(T_{env}) + \sum_{k=1,\cdots,j,\cdots,N} \boldsymbol{F}_k^{(j),neq}(T_{env}, T_k). \quad (1)$$

The first term $\boldsymbol{F}^{(j),eq}(T_{env})$ is the total Casimir force due to both zero-point and thermal fluctuations acting on the $j$-th sphere in the global thermal equilibrium at $T_{env}$. The second term describes the Casimir force contributions on the *j*-th sphere that arise in thermal nonequilibrium. $\boldsymbol{F}_k^{(j),th}(T_k)$ is the thermal Casimir force acting on the *j*-th sphere due to the thermal fluctuating currents in the *k*-th sphere, when the temperature of the *k*-th sphere is $T_k$. $\boldsymbol{F}_k^{(j),th}(T_{env})$ describes the same force when the temperature of the *k*-th sphere is at the environment temperature. The difference between the two, $\boldsymbol{F}_k^{(j),neq}(T_{env}, T_k) = \boldsymbol{F}_k^{(j),th}(T_k) - \boldsymbol{F}_k^{(j),th}(T_{env})$, only arises in thermal nonequilibrium between the *k*-th sphere and the environment as net nonequilibrium contribution of thermal Casimir force on the *j*-the sphere due the *k*-th sphere.

We clarify our terminology on Casimir and thermal Casimir forces used in the following sections. Writing out the total Casimir force as in Eq. (1) separates contributions that exist in thermal equilibrium from those that arise in thermal nonequilibrium. However, Eq. (1) does not separate zero-point and thermal contributions of total Casimir forces. In fact, the first term in Eq. (1) involves both Casimir and thermal Casimir forces when $T_{env} \neq 0$ K. Thus, we hereafter refer to the first term in Eq. (1) $\boldsymbol{F}^{(j),eq}(T_{env})$ as equilibrium Casimir and the second terms in Eq. (1) $\boldsymbol{F}_k^{(j),neq}(T_{env}, T_k)$ as nonequilibrium Casimir forces, respectively, as used in literature [59,60]. In the following sections, we introduce the formalisms of the equilibrium and nonequilibrium Casimir forces. The equilibrium Casimir force $\boldsymbol{F}^{(j),eq}(T_{env})$ is computed from the spatial derivative of the zero-point and thermal Casimir energy. The total Casimir energy, expressed as the free energy of fluctuating electromagnetic fields, is formulated in the path integral formalism and the scattering formalism [55]. The thermal Casimir force $\boldsymbol{F}_k^{(j),th}(T_k)$ is formulated as the surface integral of Maxwell's stress tensor over the *j*-th sphere due to fluctuating thermal currents in the *k*-th sphere. The total electromagnetic fields due to fluctuating current sources in the *k*-th sphere are written down by the scattering formalism. We assume that each sphere in the system can be enclosed by an imaginary sphere without overlapping with imaginary spheres enclosing other spheres as shown in Fig. 1 (a). This assumption allows us to expand electromagnetic fields in the spherical wave basis. Outside the objects, electromagnetic fields can be expanded solely by transverse solutions to the Helmholtz equation and each scattering channel in the spherical wave basis is specified by three labels $(P, l, m)$, where $P = N, M$ correspond to the TM and TE polarizations, respectively, and $(l, m)$ are the angular momentum quantum numbers of spherical harmonics. Linearly independent and complete sets of transverse solutions that have no singularity (regular waves) and satisfy the outgoing boundary condition (outgoing waves) are used to expand electromagnetic fields (see Appendices A and B for details).

## B. Equilibrium Casimir force

In this section, we briefly review part of the scattering formalism of equilibrium Casimir force that is necessary for our system, and the complete, detailed, and more general derivations can be found in [55]. In this formalism, the free energy of the fluctuating electromagnetic fields in the system comprised of the $N$ objects is formulated in the path integral formalism. The sources of the fluctuating electromagnetic fields exist both inside and outside the objects. Inside the objects, the free energy of electromagnetic fields is altered from those outside the objects due to light-matter interactions described by their dielectric and permeability tensors. We are only concerned with the part of the free energy that depends on the relative orientations and separations of the objects. This is because only the energy difference from some reference state is meaningful, and the reference state is usually taken as the objects being infinitely far apart. As a result, the part of the free energy that we are concerned with is traced back to fluctuating electromagnetic fields inside objects, which originate from fluctuating currents and charges inside the objects, and to interactions of these across the $N$ objects.

The Lipmann-Schwinger equation connects total fluctuating electromagnetic fields sourced by the objects to the way the objects scatter electromagnetic waves and to interactions between fluctuating currents inside different objects. The former is described by the so-called transition matrix (T-matrix) which we denote by $\mathcal{T}$, and the latter is described by the free Green's function. The free Green's function connects a basis of the electric field in one object to another basis in another object. Since the basis about one object is usually associated with a coordinate appropriate to the object, the free Green's function contains two bases defined separately for the two objects. Thus, in order to write down the free Green's function in terms of a coordinate about only one object, one of the bases has to be translated into the other. This is done by the so-called translation matrix, which we denote by $\mathcal{U}$. The final expression of the free energy is formally written in terms of the way the objects scatter electromagnetic waves encoded in the transition matrices of the individual objects, and relative orientations and separations between the objects encoded in the translation matrices. The transition matrices need to be separately derived for the specific objects in the system. We developed the transition matrix of a gyroelectric sphere by expanding the incident, scattered and internal waves about the sphere via vector spherical waves and then by matching the coefficients in the wave expansions so the boundary conditions are satisfied.

We focus on the case $T_{env} = 0$ K in this work. In this limit, the free energy of the system is from the zero-point energy of fluctuating electromagnetic fields, i.e., the Casimir energy, and is given as [55]:

$$E = \frac{\hbar c}{2\pi} \int_0^\infty d\kappa \log \det(\bar{M} \bar{M}_\infty^{-1}), \qquad (2)$$

where $\kappa = -i\omega/c$ is the wavenumber defined in the imaginary frequency domain, and $c$ is the speed of light in vacuum. For systems in global thermal equilibrium at finite temperatures, the integral over the imaginary wavenumber is replaced by summation over bosonic Matsubara frequency [55]. The matrix $\frac{\hbar}{2\pi} \log \det \bar{M}$ describes the spectral Casimir energy when the $N$ objects are located in the configuration of interest and the matrix $\bar{M}$ is:

$$\bar{M} \equiv \begin{bmatrix} \mathcal{T}_1^{-1} & -\mathcal{U}^{21} & \cdots & -\mathcal{U}^{N1} \\ -\mathcal{U}^{12} & \mathcal{T}_2^{-1} & \ddots & \vdots \\ \vdots & \ddots & \ddots & -\mathcal{U}^{N(N-1)} \\ -\mathcal{U}^{1N} & \cdots & -\mathcal{U}^{(N-1)N} & \mathcal{T}_N^{-1} \end{bmatrix}. \qquad (3)$$

$\frac{\hbar}{2\pi} \log \det \bar{M}_\infty^{-1}$ where $\bar{M}_\infty^{-1} = \text{diag}(\mathcal{T}_1, \cdots, \mathcal{T}_N)$ is a block diagonal matrix describes the spectral Casimir energy when the objects are infinitely far apart and thus the interactions between objects are negligible. $\mathcal{T}_i$ ($i = 1 \cdots N$) is the T-matrix of the $i$-th object. In Eq. (2), $\hbar c \kappa/2$ is equivalent to the zero-point energy of a photon at frequency $\omega$, and $\log \det(\bar{M} \bar{M}_\infty^{-1})$, which originates from the logarithm of the partition function of the system, describes the number of electromagnetic states. We can see from $\log \det(\bar{M} \bar{M}_\infty^{-1}) = \log \det \bar{M} - \log \det \bar{M}_\infty$ that the zero-point energy difference from the reference state contributes to the Casimir energy in Eq. (2).

The translation matrix $\mathcal{U}^{ji}$ translates the outgoing wave basis with respect to the spherical coordinate defined at the center of the $i$-th sphere into the regular wave basis with respect to the spherical coordinate defined at the center of the $j$-th sphere as:

$$\boldsymbol{E}_{\nu}^{out}(\kappa,\boldsymbol{r}_i) = \sum_{\nu'} \mathcal{U}_{\nu',\nu}^{ji}(\kappa,\boldsymbol{R}_{ji})\boldsymbol{E}_{\nu'}^{reg}(\kappa,\boldsymbol{r}_j), \qquad (4)$$

where $\nu = (P, l, m)$ is the label of the spherical wave basis where $P = N$ or $M$ refers to TM and TE polarizations, respectively, and $(l, m)$ are the labels of spherical harmonics. We introduce the shorthand notation $\sum_\nu \equiv \sum_{P=N,M}\sum_{l=1}^{\infty}\sum_{m=-l}^{l}$ for compactness. The spherical wave basis in the imaginary frequency used in this work is described in Appendix A. In Eq. (4), $\boldsymbol{r}_i$ and $\boldsymbol{r}_j$ refer to the same point in space measured from the origins of the coordinates defined with respect to the $i$-th and $j$-the spheres, respectively, and $\boldsymbol{R}_{ji} = \boldsymbol{r}_j - \boldsymbol{r}_i$ is the vector from the center of $i$-th sphere to that of $j$-th sphere. A schematic of this relation is shown in Fig.2 (a). The translation matrices depend only on the orientations of and separations between the objects, but do not depend on objects' material properties. The explicit expressions of the matrix elements of $\mathcal{U}^{ji}(\kappa,\boldsymbol{R}_{ji})$ derived in [55] are given in Appendix A. The transition matrix of the $i$-th sphere in the real-space representation is given as:

$$(\mathcal{T}_i)_{\nu,\nu'} = -\int d\boldsymbol{r}d\boldsymbol{r}'\, C_\nu(\kappa)\left[\boldsymbol{E}_\nu^{reg}(\boldsymbol{r},\kappa)\right]^\dagger \hat{\mathcal{T}}_i(\boldsymbol{r},\boldsymbol{r}',ic\kappa)\boldsymbol{E}_{\nu'}^{reg}(\boldsymbol{r}',\kappa), \qquad (5)$$

where the coefficients are $C_{(M,l,m)}(\kappa) = -C_{(N,l,m)}(\kappa) = \kappa$ and $\hat{\mathcal{T}}_i(\boldsymbol{r},\boldsymbol{r}',ic\kappa)$ is the T-matrix operator of the $i$-th object in the position basis representation in the imaginary frequency domain. † denotes the conjugate transpose. More concretely, suppose that an incident electric field on the $i$-th sphere is expanded in the spherical wave basis with weighted coefficients in a vector form $\boldsymbol{a}_{inc}$ and similarly those of the scattered field in a vector form $\boldsymbol{a}_{scat}$. Then the T-matrix connects these two vectors via $\boldsymbol{a}_{scat} = \mathcal{T}\boldsymbol{a}_{inc}$ as schematically shown in Fig. 2 (b). We developed the T-matrix of a gyroelectric sphere in the imaginary frequency domain by following similar procedures in the real frequency domain [61].

In this work, we are particularly interested in two-sphere and three-sphere systems, which reveal some symmetry-related physics and can serve as elementary building blocks of more complex many-body systems. For the two-sphere system, the Casimir energy Eq. (2) is simplified as:

$$E = \frac{\hbar c}{2\pi} \int_0^\infty d\kappa \log \det(I - \mathcal{T}_1 \mathcal{U}^{12} \mathcal{T}_2 \mathcal{U}^{21}), \tag{6}$$

and for the three-sphere system, the Casimir energy is:

$$E = \frac{\hbar c}{2\pi} \int_0^\infty d\kappa \log[\det(I - \mathcal{U}^{13}\mathcal{T}_3\mathcal{U}^{31}\mathcal{T}_1) \det Z], \tag{7}$$

where the matrix $Z$ is:

$$Z = I - \mathcal{U}^{23}\mathcal{T}_3\mathcal{U}^{32}\mathcal{T}_2 - (\mathcal{U}^{21}\mathcal{T}_1 + \mathcal{U}^{23}\mathcal{T}_3\mathcal{U}^{31}\mathcal{T}_1) \frac{1}{I - \mathcal{U}^{13}\mathcal{T}_3\mathcal{U}^{31}\mathcal{T}_1} (\mathcal{U}^{12}\mathcal{T}_2 + \mathcal{U}^{13}\mathcal{T}_3\mathcal{U}^{32}\mathcal{T}_2). \tag{8}$$

**C. Nonequilibrium Casimir force**

We briefly review the approaches developed to date for calculating nonequilibrium Casimir forces that can be largely classified into four groups. The first approach is based on the scattering formalism [59,60,62,63], similar to the formalism of equilibrium Casimir force [55], with the electromagnetic fields written down in the real frequency domain instead of imaginary frequency. The second approach is the Green's function method [64], which is in general applicable to calculate the linear momentum transfer, i.e., van der Waals and Casimir forces between an arbitrary number of objects. However, the compact expression based on the transmission coefficient was only given for heat transfer [64]. Both the scattering formalism and the Green's function method developed for Casimir forces in many-body systems, however, are only applicable for reciprocal systems since they assume the symmetry of Green's function operator and T-matrix operator, i.e., $\hat{G}(\boldsymbol{r},\boldsymbol{r}') = \hat{G}^T(\boldsymbol{r}',\boldsymbol{r})$ and $\hat{\mathcal{T}}(\boldsymbol{r},\boldsymbol{r}') = \hat{\mathcal{T}}^T(\boldsymbol{r}',\boldsymbol{r})$, which is only valid for reciprocal systems.

Since computations of scattering matrix and Green's function require the expansion of fields in specific basis such as plane wave, cylindrical wave, and spherical wave basis, the shape of involved objects is often restricted to highly symmetrical ones. Alternatively, the third approach based on treating volume currents in an object as unknowns does not require the expansion in specific basis, and Casimir force in complex geometries can be evaluated [65]. Lastly, the thermal discrete dipole approximation method that was previously applied for radiative heat transfer for reciprocal media and magneto-optical media [66] was recently extended for computing angular momentum transfer between two nonreciprocal nanoscale cubes [67]. While these methods can be applied to arbitrary geometries and also are asymptotically exact as volume discretization becomes finer, the computation cost increases accordingly. A computationally-efficient formalism for exactly computing nonequilibrium Casimir forces in many-body nonreciprocal systems is still desired. Based on this motivation, we develop compact formulas of thermal Casimir forces between arbitrary number of spheres with arbitrary dielectric function tensors based on the scattering formalism and the fluctuational electrodynamics. Our formalism is based on the integration of Maxwell's stress tensor over a sphere to determine thermal Casimir force on it. The total electric and magnetic fields on the surface of the sphere in the *N*-sphere system are determined by the scattering formalism, and the sources of these thermal electromagnetic fields are connected to the temperatures and material properties of the spheres by the fluctuation-dissipation theorem. While the scattering formalism of thermal Casimir forces was developed previously [59,60], it is only applicable to systems obeying Lorentz reciprocity. Moreover, the thermal Casimir force in *N*-body system is written down by using the composite T-matrix operator of the collection of bodies, which is not trivial to determine. The scattering formalism was recently extended to two nonreciprocal objects [56]. Our compact formulas can be applied to an arbitrary number of spheres with arbitrary dielectric function tensors, and are written down by the T-matrix operator of individual spheres.

We consider thermal nonequilibrium situations where the temperatures of the *N* spheres $T_j$ ($j = 1, \cdots, N$) and the environment $T_{env}$ are independently different. We assume local thermal equilibrium within each sphere by which the current fluctuations in each object can be treated independently and we do not consider inhomogeneity of temperature inside the spheres. This assumption allows us to use the fluctuation-dissipation theorem for fluctuating currents in each sphere.

In order to determine the force acting on a sphere, we take the approach of integrating the Maxwell's stress tensor over its surface. The Maxwell stress tensor is given as:

$$\sigma_{ij}(\boldsymbol{r},t) = \frac{1}{4}\varepsilon_0 \int_{-\infty}^{\infty} d\omega \left[ \begin{array}{c} \langle E_i(\boldsymbol{r},\omega)E_j^*(\boldsymbol{r},\omega)\rangle + c^2\langle B_i(\boldsymbol{r},\omega)B_j^*(\boldsymbol{r},\omega)\rangle \\ -\frac{1}{2}\langle |E|^2 + c^2|B|^2\rangle \delta_{ij} \end{array} \right], \quad (9)$$

where $i, j$ are spatial coordinates, $\varepsilon_0$ is the vacuum permittivity, and the bracket $\langle\ \rangle$ means the symmetrized ensemble average with respect to the appropriate system Hamiltonian. We defined the Fourier transform of a field $A(t)$ as $A(\omega) = \frac{1}{\pi}\int_{-\infty}^{\infty} dt e^{i\omega t} A(t)$ and $A(t) = \frac{1}{2}\int_{-\infty}^{\infty} dt e^{-i\omega t} A(\omega)$. In the following, the explicit dependence on the frequency is omitted and we write the electromagnetic fields as $E_i(\boldsymbol{r})$ and $B_i(\boldsymbol{r})$ for simplicity. The force $\boldsymbol{F}$ is given as the surface integral of the Maxwell's stress tensor over a sphere:

$$\boldsymbol{F} = \operatorname{Re} \oint_\Sigma d\boldsymbol{S} \cdot \bar{\bar{\sigma}}(\boldsymbol{r}), \tag{10}$$

where $\Sigma$ is the surface area of the sphere. Our aim is to obtain the force $\boldsymbol{F}$ in the Cartesian coordinates fixed in space, while it is convenient to perform the surface integral in the spherical coordinates when the object is a sphere. The force expressed in the spherical coordinates is:

$$\boldsymbol{F} = \operatorname{Re} \oint_\Sigma d S \begin{bmatrix} \sin\theta \cos\phi\, \sigma_{rr} + \cos\theta \cos\phi\, \sigma_{\theta r} - \sin\phi\, \sigma_{\phi r} \\ \sin\theta \sin\phi\, \sigma_{rr} + \cos\theta \sin\phi\, \sigma_{\theta r} + \cos\phi\, \sigma_{\phi r} \\ \cos\theta\, \sigma_{rr} - \sin\theta\, \sigma_{\theta r} \end{bmatrix}, \tag{11}$$

where $\oint_\Sigma dS = a^2 \int_{\theta=0}^{\theta=\pi}\int_{\phi=0}^{\phi=2\pi} \sin\theta\, d\theta d\phi$ and $a$ is the radius of the sphere.

The total electric and magnetic fields on the surface of a sphere in the $N$-sphere system are necessary in order to evaluate the surface integral Eq. (11). To determine those, we follow the derivation based on the scattering formalism given in [20] and briefly review here. Consider an electric field at a point outside the $j$-th sphere but inside the imaginary sphere that only encloses the $j$-th sphere. We call it the $j$-th region (see Fig.1 (a)). We assume this electric field is due to the fluctuating current sources inside the $k$-th sphere. Such electric field can be expanded most conveniently by the outgoing spherical wave basis from all the spheres:

$$\boldsymbol{E} = \sum_{\nu, m \neq j} d_{mk,\nu} \boldsymbol{E}_{m,\nu}^{out} + \sum_{\nu} d_{jk,\nu} \boldsymbol{E}_{j,\nu}^{out}, \tag{12}$$

where $\nu = (P, l, m)$ is the label of the spherical wave basis. We use the shorthand notation

$\sum_\nu \equiv \sum_{P=N,M} \sum_{l=1}^\infty \sum_{m=-l}^l$. The explicit expressions of the basis functions can be found in Appendix B. $\boldsymbol{E}_{m,\nu}^{out}$ is the outgoing wave basis with label $\nu$ based on the coordinates centered at the $m$-th sphere, which we call the $m$-th coordinates. The subscript of the expansion coefficient $d_{mk,\nu}$ means that it is associated with the electric field with label $\nu$ measured in the $m$-th coordinates due to the fluctuating currents inside the $k$-th sphere. The first term in Eq. (12) can be interpreted as incident field on the $j$-th sphere which we denote $\boldsymbol{E}_0 = \sum_{\nu, m \neq j} d_{mk,\nu} \boldsymbol{E}_{m,\nu}^{out}$. In order to evaluate the force acting on the $j$-th sphere, we rewrite this first term in terms of the waves expanded in the $j$-th coordinates so that we can compute the integral in Eq. (11). In fact, such incoming waves at a point in the $j$-th region can be expanded by the regular wave basis with respect to the $j$-th coordinates. Thus, we can also write down $\boldsymbol{E}_0$ as $\boldsymbol{E}_0 = \sum_j c_{jk,\nu} \boldsymbol{E}_{j,\nu}^{reg}$, where $c_{jk,\nu}$ is the expansion coefficient.

The electric field at a point expanded in the $j$-th coordinates $\boldsymbol{E}_{j,\nu}^{reg}$ and in the $m$-th coordinates $\boldsymbol{E}_{m,\nu}^{out}$ are related by the vector addition theorem, similar to Eq. (4) (see Fig. 2 (a) for schematic presentation):

$$\boldsymbol{E}_{m,\nu}^{out}(\boldsymbol{r}_m) = \sum_{\nu'} \mathcal{U}_{\nu'\nu}^{jm}(\boldsymbol{R}_{jm}) \boldsymbol{E}_{j,\nu'}^{reg}(\boldsymbol{r}_j), \tag{13}$$

where $\mathcal{U}_{\nu'\nu}^{jm}$ is the matrix elements of the translation matrix $\mathcal{U}^{jm}$ between the $j$-th and $m$-th coordinates, and $\boldsymbol{R}_{jm} = \boldsymbol{r}_j - \boldsymbol{r}_m$. We then find the relation between the expansion coefficients in the vector form as:

$$\boldsymbol{c}_{jk} = \sum_{m \neq j} \mathcal{U}^{jm} \boldsymbol{d}_{mk}. \tag{14}$$

The second term in Eq. (12) can be interpreted as the field scattered by the $j$-th sphere when $j \neq k$. When $j = k$, the emitted fields need to be added to the second term. We first consider the case $j \neq k$. The Lippmann-Schwinger equation [68] for a point inside the $j$-th region is

$$\boldsymbol{E}(\boldsymbol{r}) = \boldsymbol{E}_0(\boldsymbol{r}) + \int d\boldsymbol{r}' d\boldsymbol{r}'' G_0(\boldsymbol{r},\boldsymbol{r}') \hat{\mathcal{T}}_j(\boldsymbol{r}',\boldsymbol{r}'') \boldsymbol{E}_0(\boldsymbol{r}''), \tag{15}$$

where $\hat{\mathcal{T}}_j(\boldsymbol{r}', \boldsymbol{r}'')$ is the T-matrix operator about the $j$-th sphere in the position basis representations and $G_0(\boldsymbol{r}, \boldsymbol{r}')$ is the free Green's function expanded in the spherical wave basis for $|\boldsymbol{r}| > |\boldsymbol{r}'|$ as [59]:

$$G_0(\boldsymbol{r}, \boldsymbol{r}') = i \sum_v \boldsymbol{E}_v^{out}(\boldsymbol{r}) \otimes \boldsymbol{E}_{\sigma(v)}^{reg}(\boldsymbol{r}'), \tag{16}$$

where $\sigma(v) \equiv (P, l, -m)$ if $v = (P, l, m)$ and the vector product $(\boldsymbol{a} \otimes \boldsymbol{b})_{ij} \equiv a_i b_j$ is used. By comparing the second terms of Eqs. (12) and (15) and using Eq. (16), we find the relation between the expansion coefficients for incident and scattered waves as:

$$d_{jk,v} \equiv d_{jk,v}^{scat} = \sum_{v'} \mathcal{T}_{j,vv'} c_{jk,v'} \text{ for } j \neq k, \tag{17}$$

where the matrix elements of the $T$-matrix of the $j$-th sphere are defined as [59]:

$$(\mathcal{T}_j)_{vv'} = i \int d\boldsymbol{r} d\boldsymbol{r}' \left[\boldsymbol{E}_{\sigma(v)}^{reg}(\boldsymbol{r})\right]^T \hat{\mathcal{T}}_j(\boldsymbol{r}, \boldsymbol{r}') \boldsymbol{E}_{v'}^{reg}(\boldsymbol{r}'). \tag{18}$$

In the case where $j = k$, the first term in Eq. (12) remains the same. However, the second term has contribution from the emission denoted by the expansion coefficient $d_{k,v}^0$ in addition to the scattered waves as given in Eq. (17). As a result, we can write the expansion coefficients for outgoing waves for the case $j = k$ as

$$d_{kk,v} = d_{k,v}^0 + d_{kk,v}^{scat} \text{ for } j = k. \tag{19}$$

Combining Eqs. (14), (17) and (19), we can determine the expansion coefficients $d_{jk,v}$ as:

$$\begin{bmatrix} \boldsymbol{d}_{1k} \\ \vdots \\ \boldsymbol{d}_{kk} \\ \vdots \\ \boldsymbol{d}_{Nk} \end{bmatrix} = (\mathcal{I} - \bar{\mathcal{T}}\bar{\mathcal{U}})^{-1} \begin{bmatrix} 0 \\ \vdots \\ \boldsymbol{d}_k^0 \\ \vdots \\ 0 \end{bmatrix}, \tag{20}$$

for $j = 1, \cdots, N$. The matrix $\mathcal{I}$ is the identity matrix. The matrices $\bar{\mathcal{T}}$ and $\bar{\mathcal{U}}$ are defined as:

$$\bar{\mathcal{T}} \equiv \begin{bmatrix} \mathcal{T}_1 & 0 & \cdots & 0 \\ 0 & \mathcal{T}_2 & \ddots & \vdots \\ \vdots & \ddots & \ddots & 0 \\ 0 & \cdots & 0 & \mathcal{T}_N \end{bmatrix}, \tag{21}$$

and

$$\bar{\mathcal{U}} \equiv \begin{bmatrix} 0 & \mathcal{U}^{12} & \cdots & 0 \\ \mathcal{U}^{21} & 0 & \ddots & \vdots \\ \vdots & \ddots & \ddots & \mathcal{U}^{(N-1)N} \\ \mathcal{U}^{N1} & \cdots & \mathcal{U}^{N(N-1)} & 0 \end{bmatrix}. \tag{22}$$

Although we use the letters $\mathcal{T}$ and $\mathcal{U}$ to denote the transition and translation matrices both in the real and imaginary frequency domains, we distinguish them from the context. Note that $\boldsymbol{d}_{jk}$ in Eq. (20) is the column vector of the size $2n_d \equiv 2l_{max}(l_{max} + 2)$ where $l_{max}$ is the cutoff number of the quantum number $l$ of spherical harmonics. Thus, the size of the matrices $\mathcal{T}_j$ and $\mathcal{U}^{ij}$ is $2n_d \times 2n_d$. By defining $\bar{\mathcal{Q}} \equiv (\mathcal{I} - \bar{\mathcal{T}}\bar{\mathcal{U}})^{-1}$, the $(j,k)$-th block element of the matrix $\bar{\mathcal{Q}}_{jk}$ relates the outgoing waves from the body $j$ and the emitted waves from the body $k$. The expansion coefficients $c_{jk}$ can be found by using Eq. (14) as:

$$\begin{bmatrix} \boldsymbol{c}_{1k} \\ \vdots \\ \boldsymbol{c}_{kk} \\ \vdots \\ \boldsymbol{c}_{Nk} \end{bmatrix} = \bar{\mathcal{W}} \begin{bmatrix} 0 \\ \vdots \\ \boldsymbol{d}_k^0 \\ \vdots \\ 0 \end{bmatrix}, \tag{23}$$

where $\bar{\mathcal{W}} \equiv \bar{\mathcal{U}}(\mathcal{I} - \bar{\mathcal{T}}\bar{\mathcal{U}})^{-1}$. Finally, we can write the total electric field in the $j$-th region in the $N$-sphere system as:

$$\boldsymbol{E}(\boldsymbol{r}) = \sum_{\nu} \left[ d_{jk,\nu} \boldsymbol{E}_{j,\nu}^{out}(\boldsymbol{r}) + c_{jk,\nu} \boldsymbol{E}_{j,\nu}^{reg}(\boldsymbol{r}) \right]. \tag{24}$$

Using the Faraday's law $i\omega \boldsymbol{B}(\boldsymbol{r}) = \nabla \times \boldsymbol{E}(\boldsymbol{r})$ and the identities for the spherical wave basis (see Eq. (B3) in Appendix B), the total magnetic field is written as:

$$\boldsymbol{B} = \frac{k_0}{i\omega} \sum_{\nu} \left[ d_{jk,\nu} \boldsymbol{E}_{j,\bar{\nu}}^{out} + c_{jk,\nu} \boldsymbol{E}_{j,\bar{\nu}}^{reg} \right], \tag{25}$$

where $k_0 = \omega/c$ is the wavenumber in vacuum, $c$ is the speed of light in vacuum and the bar on the label switches the polarization, i.e., $\bar{\nu} \equiv (M, l, m)$ if $\nu = (N, l, m)$ and vice versa.

We can obtain the thermal Casimir force acting on a sphere by inserting the total electric and magnetic fields in the Maxwell's stress tensor and integrating over the surface of the sphere as in Eq. (11). We illustrate the evaluation of such integrals by showing the derivation of a single term in the $z$-component of the force. The first term of the $z$-component of the force requires the evaluation of the integral of the form $\text{Re} \oint_\Sigma dS \cos\theta\, \sigma_{rr}$. In this term, there exists three contributions as given by Eq. (11). We only show the derivation of the term $\text{Re} \oint_\Sigma dS \cos\theta\, \langle E_r E_r^* \rangle$ and other terms can be evaluated similarly. By expanding this term using Eq. (24), we have

$$\text{Re} \oint_\Sigma dS \cos\theta \sum_{\nu,\nu'} \langle [d_{jk,\nu} E_{j,\nu,r}^{out} + c_{jk,\nu} E_{j,\nu,r}^{reg}] [d_{jk,\nu'}^* E_{j,\nu',r}^{out*} + c_{jk,\nu'}^* E_{j,\nu',r}^{reg*}] \rangle, \tag{26}$$

where the subscript $r$ means the $r$-component of the electric field vector. This is non-zero only for the $N$ polarization since the $r$ component for the $M$ polarization is zero (see Appendix B). Writing the label as $\nu = (N, l, m)$ and $\nu' = (N, l', m')$, we need to evaluate the integral of the form

$$\text{Re} R_j^2 \oint d\Omega \cos\theta\, E_{j,(N,l,m),r}^\alpha E_{j,(N,l',m'),r}^{\beta*}$$
$$= \text{Re} \frac{1}{k_0} \sqrt{(-1)^m} (-1)^{m'} \sqrt{(-1)^{m'}} \sqrt{l(l+1)l'(l'+1)} \times$$
$$z_l^{(\alpha)}(k_0 R_j) z_{l'}^{(\beta)*}(k_0 R_j) \oint d\Omega \cos\theta\, Y_{lm} Y_{l'm'}^*, \tag{27}$$

where $\alpha, \beta$ are the labels for regular and outgoing wave basis, $Y_{lm}$ are the spherical harmonics, and the function $z_l^{(\alpha)}(x)$ is the spherical Bessel function when $\alpha = reg$ and the spherical Hankel function of the first kind when $\alpha = out$. $\oint d\Omega \equiv \int_0^\pi \sin\theta \, d\theta \int_0^{2\pi} d\phi$ is the surface integral of the sphere of unit radius. We use the following identity [69]:

$$\oint d\Omega \cos\theta \, Y_{lm} Y^*_{l'm'}$$
$$= \left[\delta_{l,l'+1}\sqrt{\frac{(l'+1)^2 - m'^2}{(2l'+1)(2l'+3)}} + \delta_{l,l'-1}\sqrt{\frac{l'^2 - m'^2}{(2l'-1)(2l'+1)}}\right]\delta_{m,m'}. \tag{28}$$

Then if we define the function

$$\mathcal{A}^{(\alpha,\beta)}_{(N,l,m),(N,l',m')}(x) = z_l^{(\alpha)}(x) z_{l'}^{(\beta)*}(x) \sqrt{(-1)^m}(-1)^{m'}\sqrt{(-1)^{m'}}\sqrt{l(l+1)l'(l'+1)} \times$$
$$\left[\delta_{l,l'+1}\sqrt{\frac{(l'+1)^2 - m'^2}{(2l'+1)(2l'+3)}} + \delta_{l,l'-1}\sqrt{\frac{l'^2 - m'^2}{(2l'-1)(2l'+1)}}\right]\delta_{m,m'}, \tag{29}$$

Eq. (26) can be written down as:

$$\sum_{l,m}\sum_{l'm'} \frac{1}{k_0} \text{Re} \begin{bmatrix} \langle d_{jk,(N,l,m)} d^*_{jk,(N,l',m')}\rangle \mathcal{A}^{(out,out)}_{(N,l,m),(N,l',m')}(k_0 R_j) \\ + \langle d_{jk,(N,l,m)} c^*_{jk,(N,l',m')}\rangle \mathcal{A}^{(out,reg)}_{(N,l,m),(N,l',m')}(k_0 R_j) \\ + \langle c_{jk,(N,l,m)} d^*_{jk,(N,l',m')}\rangle \mathcal{A}^{(reg,out)}_{(N,l,m),(N,l',m')}(k_0 R_j) \\ + \langle c_{jk,(N,l,m)} c^*_{jk,(N,l',m')}\rangle \mathcal{A}^{(reg,reg)}_{(N,l,m),(N,l',m')}(k_0 R_j) \end{bmatrix}, \tag{30}$$

where the shorthand notation $\Sigma_{l,m} = \sum_{l=1}^{\infty}\sum_{m=-l}^{l}$ is used for the summation. All four terms of the ensemble average of the expansion coefficients can be written down in terms of the emission coefficients $\langle d^0_{kv} d^{0*}_{kv'}\rangle$ from the relations we obtained in Eqs. (20) and (23). The fluctuation-dissipation theorem dictates that this ensemble average is related to the so-called radiation operator about the $k$-th sphere, $\mathcal{R}_k$ [59] as:

$$\langle \boldsymbol{d}_k^0 \boldsymbol{d}_k^{0\dagger} \rangle = \frac{4}{\pi} \omega \Theta(\omega, T_k) \mu_0 \mathcal{R}_k, \tag{31}$$

where $\Theta(\omega, T) = \hbar\omega \left(e^{\frac{\hbar\omega}{k_B T}} - 1\right)^{-1}$ and $\mu_0$ is the magnetic permeability of vacuum. Although the fluctuation-dissipation theorem based on the symmetrized correlation function includes the zero-point energy term in $\Theta(\omega, T)$, this does not appear in nonequilibrium contributions of Casimir force $\boldsymbol{F}_k^{(j),neq}(T_{env}, T_k)$ since we are only concerned with the difference $\boldsymbol{F}_k^{(j),neq}(T_{env}, T_k) = \boldsymbol{F}_k^{(j),th}(T_k) - \boldsymbol{F}_k^{(j),th}(T_{env})$ in Eq. (1) and the zero-point energy contribution cancels out. In the spherical wave basis, the radiation operator is given as [20,59]:

$$\mathcal{R}_k = -\frac{\mathcal{T}_k + \mathcal{T}_k^\dagger}{2} - \mathcal{T}_k \mathcal{T}_k^\dagger. \tag{32}$$

We now evaluated Eq. (30) completely. Other terms in the $z$-component force can be evaluated similarly by using the identities listed in Appendix B. As a result, the formula for the $z$-component thermal Casimir force on the $j$-th sphere due to thermal emission from the $k$-th sphere can be obtained as:

$$\left(F_k^{(j),th}\right)_z = \frac{2}{\pi c} \int_0^\infty d\omega\, \Theta(\omega, T_k) \operatorname{ReTr} \begin{bmatrix} \bar{\mathcal{Q}}_{jk} \mathcal{R}_k \bar{\mathcal{Q}}_{jk}^\dagger \mathcal{P}_z^{(out,out),T}(k_0 R_j) \\ +\bar{\mathcal{Q}}_{jk} \mathcal{R}_k \bar{\mathcal{W}}_{jk}^\dagger \mathcal{P}_z^{(out,reg),T}(k_0 R_j) \\ +\bar{\mathcal{W}}_{jk} \mathcal{R}_k \bar{\mathcal{W}}_{jk}^\dagger \mathcal{P}_z^{(reg,out),T}(k_0 R_j) \\ +\bar{\mathcal{W}}_{jk} \mathcal{R}_k \bar{\mathcal{W}}_{jk}^\dagger \mathcal{P}_z^{(reg,reg),T}(k_0 R_j) \end{bmatrix}, \tag{33}$$

where $T$ denotes the transpose. The matrix $\mathcal{P}_z^{(\alpha,\beta)}$ ($\alpha, \beta$ = reg, out) is defined as:

$$\mathcal{P}_z^{(\alpha,\beta)} = \mathcal{A}^{(\alpha,\beta)} - \frac{\mathcal{B}^{(\alpha,\beta)} + \bar{\mathcal{B}}^{(\alpha,\beta)}}{2} - \mathcal{C}^{(\alpha,\beta)}, \tag{34}$$

where the matrix $\mathcal{A}^{(\alpha,\beta)}$ describes the contribution from the first two terms proportional to $E_r E_r$ and $B_r B_r$ of the stress component $\sigma_{rr}$ and these two contributions form a diagonal matrix as:

$$\mathcal{A}^{(\alpha,\beta)} = \begin{bmatrix} \left[\mathcal{A}^{(\alpha,\beta)}_{(N,l,m),(N,l',m')}\right] & [0] \\ [0] & \left[\mathcal{A}^{(\alpha,\beta)}_{(M,l,m),(M,l',m')}\right] \end{bmatrix}. \tag{35}$$

Here, square brackets are used to denote that $\left[\mathcal{A}^{(\alpha,\beta)}_{(N,l,m),(N,l',m')}\right]$ is a $n_d$ by $n_d$ matrix with its matrix element given by $\mathcal{A}^{(\alpha,\beta)}_{(N,l,m),(N,l',m')}$. The matrices $\mathcal{B}^{(\alpha,\beta)}$ and $\bar{\mathcal{B}}^{(\alpha,\beta)}$ describe the contributions from the dot products of the electric and magnetic fields in the stress tensor component proportional to $\sigma_{rr}$, respectively, and $\mathcal{C}^{(\alpha,\beta)}$ describes the contribution from the stress tensor component proportional to $\sigma_{\theta r}$. The explicit expressions for these matrices are given in Appendix B.

We can also evaluate the surface integrals for the $x$ and $y$ components of the force in Eq. (11). The $x$-component of the thermal Casimir force on the $j$-th sphere due to the source in $k$-th sphere is given as:

$$\left(F_k^{(j),th}\right)_x = \frac{2}{\pi c}\int_0^\infty d\omega\, \Theta(\omega, T_k) \text{ReTr}\begin{bmatrix} \bar{\mathcal{Q}}_{jk}\mathcal{R}_k\bar{\mathcal{Q}}_{jk}^\dagger \mathcal{P}_x^{(out,out),T}(k_0 R_j) \\ +\bar{\mathcal{Q}}_{jk}\mathcal{R}_k\bar{\mathcal{W}}_{jk}^\dagger \mathcal{P}_x^{(out,reg),T}(k_0 R_j) \\ +\bar{\mathcal{W}}_{jk}\mathcal{R}_k\bar{\mathcal{Q}}_{jk}^\dagger \mathcal{P}_x^{(reg,out),T}(k_0 R_j) \\ +\bar{\mathcal{W}}_{jk}\mathcal{R}_k\bar{\mathcal{W}}_{jk}^\dagger \mathcal{P}_x^{(reg,reg),T}(k_0 R_j) \end{bmatrix}, \tag{36}$$

where the matrix $\mathcal{P}_x^{(\alpha,\beta)}$ ($\alpha,\beta$= reg, out) is

$$\mathcal{P}_x^{(\alpha,\beta)} = \mathcal{D}^{(\alpha,\beta)} - \frac{\mathcal{E}^{(\alpha,\beta)} + \bar{\mathcal{E}}^{(\alpha,\beta)}}{2} - \mathcal{F}^{(\alpha,\beta)} - \mathcal{G}^{(\alpha,\beta)}, \tag{37}$$

and the explicit expressions of the matrix elements are given in Appendix B. Finally, the *y*-component of the thermal Casimir force on the *j*-th sphere due to the source in *k*-th sphere is given as:

$$\left(F_k^{(j),th}\right)_y = \frac{2}{\pi c}\int_0^\infty d\omega\,\Theta(\omega,T_k)\text{ReTr}\begin{bmatrix} \bar{\mathcal{Q}}_{jk}\mathcal{R}_k\bar{\mathcal{Q}}_{jk}^\dagger\mathcal{P}_y^{(out,out),T}(k_0R_j) \\ +\bar{\mathcal{Q}}_{jk}\mathcal{R}_k\bar{\mathcal{W}}_{jk}^\dagger\mathcal{P}_y^{(out,reg),T}(k_0R_j) \\ +\bar{\mathcal{W}}_{jk}\mathcal{R}_k\bar{\mathcal{Q}}_{jk}^\dagger\mathcal{P}_y^{(reg,out),T}(k_0R_j) \\ +\bar{\mathcal{W}}_{jk}\mathcal{R}_k\bar{\mathcal{W}}_{jk}^\dagger\mathcal{P}_y^{(reg,reg),T}(k_0R_j) \end{bmatrix}, \quad (38)$$

where the matrix $\mathcal{P}_y^{(\alpha,\beta)}$ ($\alpha,\beta$= reg, out) is

$$\mathcal{P}_y^{(\alpha,\beta)} = \mathcal{H}^{(\alpha,\beta)} - \frac{\mathcal{J}^{(\alpha,\beta)} + \bar{\mathcal{J}}^{(\alpha,\beta)}}{2} + \mathcal{K}^{(\alpha,\beta)} + \mathcal{L}^{(\alpha,\beta)}, \quad (39)$$

and the explicit expressions of the matrix elements are given in Appendix B. With these expressions, we are able to calculate the thermal Casimir force on the *j*-th sphere due to the fluctuating current sources in the *k*-th sphere in the presence of the other spheres. By expanding the fields with large enough cutoff of the polynomial degree of spherical harmonics, $l_{max}$, these formulas in principle give exact results for the thermal Casimir force.

**D. Material properties**

In this work, we consider magnetic Weyl semimetals and silica as representative materials that exhibit nonreciprocal and reciprocal electromagnetic responses, respectively. The developed formalism is more general and is directly applicable to modeling any form of dielectric function tensors. The dielectric function model for magnetic Weyl semimetals is based on the linear-response formula of electrical conductivity under the effective low-energy Hamiltonian describing Weyl fermions $H_\eta = \hbar v_F \boldsymbol{\sigma} \cdot (\eta \boldsymbol{k} + \boldsymbol{b})$, where $\boldsymbol{\sigma}$ is the Pauli matrix vector and $\eta$ is the chirality of a Weyl node. As a prototypical model, we consider a magnetic Weyl semimetal that exhibits two Weyl nodes with opposite chirality at the same energy, which is the minimum number of the Weyl nodes in time-reversal-symmetry breaking Weyl semimetals. We assume the Weyl node separation of $2|\boldsymbol{b}|$ in the z-direction in the momentum space as shown in Fig. 1 (b). The dielectric function tensor of the magnetic Weyl semimetal can be expressed as:

$$\bar{\bar{\varepsilon}}(\omega) = \begin{bmatrix} \varepsilon_d & i\varepsilon_{AHE} & 0 \\ -i\varepsilon_{AHE} & \varepsilon_d & 0 \\ 0 & 0 & \varepsilon_d \end{bmatrix}, \tag{40}$$

where the off-diagonal element $\varepsilon_{AHE} = be^2/(2\pi^2\varepsilon_0\hbar\omega)$ arises due to the intrinsic mechanism of the anomalous Hall effect [70]. For the diagonal elements, the Kubo-Greenwood formula under the random phase approximation gives [71]:

$$\varepsilon_d(\omega) = \varepsilon_\infty + \frac{ir_s g}{6\Omega_0}\Omega G\left(\frac{\Omega}{2}\right)$$
$$-\frac{r_s g}{6\pi\Omega_0}\left[\frac{4}{\Omega}\left\{1+\frac{\pi^2}{3}\left(\frac{k_B T}{E_F}\right)^2\right\} + 8\Omega\int_0^\Lambda \frac{G(\xi) - G\left(\frac{\Omega}{2}\right)}{\Omega^2 - 4\xi^2}\xi d\xi\right], \tag{41}$$

where $\varepsilon_\infty$ is the dielectric constant in the high-frequency limit, $r_s = \frac{e^2}{(4\pi\varepsilon_0\hbar v_F)}$ is the effective fine-structure constant, $g$ is the number of Weyl nodes, and $E_F$ is the chemical potential. The non-dimensional frequencies are defined as $\Omega_0 = \hbar\omega/E_F$ and $\Omega = (\hbar\omega + i\hbar\gamma)/E_F$, where $\gamma$ is the scattering rate. $\Lambda$ is the cutoff energy normalized by the chemical potential above which the Dirac spectrum is no longer linear. $G(E) = n_F(-E) - n_F(E)$, where $n_F$ is the Fermi-Dirac distribution at temperature $T$. This expression includes both intra- and inter-band transitions of Weyl fermions. In this work, we assume a constant relative magnetic permeability $\mu = 1$, $E_F = 150\text{meV}$, $2b = 0.4\text{Å}^{-1}$, $v_F = 10^5\text{m/s}$, $\hbar\gamma = 1.5\text{meV}$, $\Lambda = 3E_F$, and $g = 2$ for the parameters of magnetic Weyl semimetals and we do not consider the Fermi-arc surface states. Since the dielectric function is analytical in the upper complex plane, the dielectric function in the imaginary frequency is obtained by replacing $\omega \to i\xi = ic\kappa$ in Eq. (41). By taking the limit $T \to 0$ limit since we consider the environment temperature $T_{env} = 0$ K, we have:

$$\varepsilon_d(i\xi) = \varepsilon_\infty + \frac{2r_s g}{3\pi}\frac{1}{\Xi_0\Xi_0} + \frac{r_s g}{6\pi}\ln\left|\frac{\Lambda^2}{\Xi_0^2 + 4}\right|, \tag{42}$$

where $\Xi_0 = \frac{\hbar \xi}{E_F}$. The background dielectric constant $\varepsilon_\infty = 6.2$ is used in this work. Another option is to use $\varepsilon_\infty = 1$, but we found the force is not significantly affected by the choice of this value.

We comment on our choice of the constant relative magnetic permeability $\mu = 1$ for magnetic Weyl semimetals. When magnetic Weyl semimetals are ferromagnets, spontaneous magnetizations in the Weyl semimetals create magnetic forces, which can be much larger than Casimir forces. Our focus in this work, however, is on nonreciprocal effects due to gyroelectricity on Casimir interactions and thus we do not consider the forces due to magnetizations. In the next section, we discuss Casimir forces between a magnetic Weyl semimetal sphere and a silica sphere as a case where the magnetic force is negligible but nonreciprocal effects can be observed. When ferromagnetic Weyl semimetals do not possess spontaneous magnetization, magnetic properties can still appear in Casimir interactions through the magnetic permeability $\mu > 1$ at low frequencies. It was predicted that the inclusion of the magnetic permeability of ferromagnets Ni and Fe in Casimir forces between two plates at 300 K contributes only through the zero-frequency term in the Lifshitz formula and can vary the Casimir force by 0.5 ~ 2 times [72]. However, we set the magnetic permeability of magnetic Weyl semimetal $\mu = 1$ in the calculation of equilibrium Casimir force and detailed study of its effects is left for future study. Lastly, although the magnetic permeability tensor of a ferromagnet is gyromagnetic [73], the relevant frequency on Casimir force $\sim c/d$, where $d$ is the inter-sphere separation, is high enough that we can consider the magnetic permeability to be isotropic.

The dielectric function of silica is taken from [74]. The dielectric function in the imaginary frequency is determined by the integral expression [75]

$$\varepsilon(i\xi) = 1 + \frac{2}{\pi} \int_0^\infty d\omega \frac{\omega \text{Im}[\varepsilon(\omega)]}{\omega^2 + \xi^2}. \tag{43}$$

### E. Computation details

In this section, we introduce the workflow of computations. To begin with, we describe the work flow of thermal Casimir force computations. We first define the system by selecting the locations, the radii, and the temperatures of the spheres. Also, we estimate the truncation number $l_{max}$ of the infinite sum over $l$. We found that $l_{max} = 7$ results in good convergence for the minimum inter-sphere separation $d = 3\mu m$ for the radii of the spheres $R = 1\mu m$ and for the sphere temperatures of 300K considered in this work. For each frequency in the frequency integrals of the thermal Casimir force expressions, we compute the dielectric function tensor of the spheres. Once they are determined, the T-matrices $\mathcal{T}_j$ of individual spheres and the translation matrices $\mathcal{U}^{ij}$ are computed.

In order to represent the expansion coefficients of electromagnetic fields in compact forms, the combined index $l(l+1) + m$ is used [76]. In this representation, a column vector of the expansion coefficients is made of two vectors for each polarization. For example, the expansion coefficients of the regular waves about the $j$-th sphere due to current fluctuations in the $k$-th sphere can be written as $\boldsymbol{c}_{jk} = \left[[c_{(N,l,m)}], [c_{(M,l,m)}]\right]^T$, where $[c_{(P,l,m)}]$ with $P = N$ or $M$ is a 1 by $l_{max}(l_{max} + 2)$ vector. The expansion coefficient for the mode labeled by $(P, l, m)$, i.e., $c_{(P,l,m)}$, is a $(l(l+1) + m)$-th element of the vector $[c_{(P,l,m)}]$. The other expansion coefficients $\boldsymbol{d}_{jk}$ and $\boldsymbol{d}_k^0$ are also expressed similarly. Using this representation, the elements of the T-matrices and translation matrices are computed. The T-matrix of a magnetic Weyl semimetal sphere in the real frequency domain is obtained by expanding the incident, scattered and internal waves about the sphere via vector spherical waves and then by matching the coefficients in the wave expansions so the electromagnetic boundary conditions at the surface of the sphere are satisfied [61,77]. To efficiently compute the matrix elements of the translation matrices, we used the recursive formula derived in [78-80]. Once the T-matrices and the translation matrices are computed we can obtain the matrices $\bar{\mathcal{Q}}_{jk}$, $\bar{\mathcal{W}}_{jk}$, and $\mathcal{R}_k$. We can also compute the matrices $\mathcal{P}_{x,y,z}^{(\alpha,\beta)}(k_0 R_j)$ for all the combinations of $\alpha, \beta = reg, out$ using the expressions derived in this work (see Appendix B). Finally, by repeating this procedure for different frequencies and performing the integration, we obtain thermal Casimir forces $\boldsymbol{F}_k^{(j),th}(T_k)$ and $\boldsymbol{F}_k^{(j),th}(T_{env})$, and the nonequilibrium Casimir force $\boldsymbol{F}_k^{(j),neq}(T_{env}, T_k) = \boldsymbol{F}_k^{(j),th}(T_k) - \boldsymbol{F}_k^{(j),th}(T_{env})$ is obtained.

Similarly, to compute the equilibrium Casimir force, the Casimir energy for each frequency is computed by computing the T-matrices and the translation matrices in the imaginary frequencies. The T-matrices in the imaginary frequency domain are obtained by following the same procedure as those in the real frequency domain. The matrix elements of the translation matrices in the imaginary frequency domain are computed by the expressions given in Appendix A. By integrating over the imaginary frequency, we obtain the Casimir energy. By computing the spatial derivative of the Casimir energy numerically by performing two separate computations of the Casimir energy with objects' locations being displaced, we obtain the equilibrium Casimir forces in the system.

## III. Results and Discussion

In this section, we discuss total Casimir forces between reciprocal and nonreciprocal spheres using the developed formalisms. In section A, we first develop an intuition about thermal Casimir force interactions by discussing thermal radiation fields in a two-sphere system where at least one of the spheres is nonreciprocal. Then we discuss total Casimir forces between two magnetic Weyl semimetal spheres and between a magnetic Weyl semimetal sphere and a silica sphere in thermal nonequilibrium. The dependence of the Casimir energy on different orientations of momentum-space Weyl node directions in the two magnetic Weyl semimetal spheres is also discussed. In section B, we discuss total Casimir forces and Casimir energy between three magnetic Weyl semimetal spheres for different orientations of momentum space Weyl node separation directions in the spheres.

### A. Two-sphere systems

Using the developed formalisms of equilibrium and nonequilibrium Casimir forces, we first study the force interactions between two magnetic Weyl semimetal spheres. As we aim to study different force interactions for different orientations of the momentum-space Weyl node separation directions in the spheres, we consider three distinct configurations of the orientations. Below, we use the term "*b* vector" to refer to the direction of momentum-space Weyl node separation in a sphere. We show that when the Weyl node separation direction is misaligned from the inter-sphere direction, the net lateral thermal Casimir force can arise in thermal nonequilibrium and the directions of the forces can change according to the symmetry of the system. When two magnetic Weyl semimetal spheres exhibit strong ferromagnetism, a force due to static magnetic moments in general overrides the Casimir force. We consider the total Casimir force between a magnetic Weyl semimetal and a silica sphere and show that the lateral thermal Casimir force still exists in a system where at least one of the spheres is nonreciprocal. In the following results, we assume the spheres have the same radius of $R = 1\mu m$ and are at 300 K. We also assume the environment is at $T_{env} = 0K$. When we discuss thermal equilibrium situations, we set the temperatures of the spheres at 0K.

First, we develop an intuitive understanding of the thermal Casimir force acting on a sphere due to thermal emission from a magnetic Weyl semimetal sphere. Figure 3 (a) and (b) show the spatial distribution of the Poynting flux vector around a magnetic Weyl semimetal sphere (blue) due to its thermal emission in the presence of the gray sphere. The gray sphere can be either reciprocal or nonreciprocal. The two panels correspond to the two different magnitudes of the **b** vectors. The **b** vector of the blue sphere points to the positive $k_z$-direction, or equivalently the gyration axis of the sphere points to the positive z-axis. The frequency-integrated Poynting flux vector on the z=0 plane is plotted and the vectors are normalized so that only their directions are meaningful. When the Weyl node separation |**b**| is non-zero, the Poynting flux vector distribution shows a "swirling" or vortex-like pattern in the near-field of the sphere, in which the Poynting vector possesses a tangential component near the surface of the sphere rather than being purely radial.

It was previously discussed that the swirling pattern of thermal emission from a magneto-optical sphere has its origin in the degeneracy lifting of the dipolar resonant modes in a sphere in the presence of an external magnetic field [27]. In the absence of the magnetic field, the three dipolar modes associated with the momentum quantum numbers *m* of spherical harmonics have the same eigenfrequency, i.e., $\omega_{l=1,m=-1} = \omega_{l=1,m=0} = \omega_{l=1,m=1}$. In the presence of an external magnetic field, however, this degeneracy between the two modes $\omega_{l=1,m=-1}$ and $\omega_{l=1,m=1}$ is lifted due to the breakdown of time-reversal symmetry, and these two modes are differently thermally populated. As a result, thermal radiation carries non-zero angular momentum in the direction of gyration, i.e., the *z*-direction.

In the case of magnetic Weyl semimetals, the net Berry curvature plays an equivalent role of the magnetic field in the momentum space. When we turn off the **b** vector, i.e., $|\boldsymbol{b}| = 0$, the thermal emission from the magnetic Weyl semimetal sphere has the radial component only, as shown in Fig. 3 (b), which is typical radiation from an isotropic reciprocal sphere. Different pattern of thermal emission from the magnetic Weyl semimetal sphere results in different radiation pressure on the gray sphere. In Fig. 3 (c) and (d), we show the Poynting flux vector distribution around the gray sphere due to thermal emission from the magnetic Weyl semimetal sphere for the two **b** vectors. When |**b**| is non-zero, it is expected that the swirling thermal emission pattern creates the

*y*-direction radiation pressure on the gray sphere, i.e., lateral thermal Casimir force. Contrarily, when the Weyl node separation is turned off, the Poynting vector distribution around the gray sphere is symmetric about the *y*-axis, creating no radiation pressure in this direction. Given that this swirling thermal emission pattern only exists in the near-field, i.e., few radii away from the sphere center, the force in the *y*-direction is expected to disappear in the far-field.

Figure 4 (a), (d), and (g) show three cases where the net lateral thermal Casimir forces arise in thermal nonequilibrium due to a non-zero gyration vector component perpendicular to the inter-sphere direction (*x*-direction). The cases (a) and (d) show two magnetic Weyl semimetal spheres where the ***b*** vectors are parallel and antiparallel, respectively, and are all perpendicular to the inter-sphere direction. For the case (a) where the ***b*** vectors in both of the spheres are parallel, the system is inversion-symmetric about the origin. Therefore, if sphere 1 experiences the force $\boldsymbol{F}^{(1)}$, then the force on sphere 2 is $\boldsymbol{F}^{(2)} = -\boldsymbol{F}^{(1)}$. This indicates that the lateral thermal Casimir forces have the opposite sign, and the two spheres will experience the thrust to rotate about the origin as schematically shown in Fig. 4 (a). Contrarily, the case (d) where the ***b*** vectors in both of the spheres are antiparallel, the system is mirror-symmetric about the *yz*-plane because the vector ***b*** is a static pseudovector and changes the direction upon mirror reflection. Thus, the force acting on sphere 1 $\boldsymbol{F}^{(1)} = (F_x^{(1)}, F_y^{(1)})$ is related to the force acting on sphere 2 as $\boldsymbol{F}^{(2)} = (-F_x^{(1)}, F_y^{(1)})$. This indicates that the lateral thermal Casimir forces have the same sign, and the two spheres will experience the thrust to propel each other in the same direction as shown in Fig. 4 (d). Therefore, we can realize two qualitatively distinct thrusts on the spheres by changing the orientations of the spheres.

We confirmed the intuitively predicted system behavior for different spheres orientation by exact numerical calculations. The total Casimir forces in the *x*-direction of the two cases (a) and (d) are shown in Fig. 4 (b) and (e), respectively. The figures show the absolute values of the forces on the left sphere (sphere 2) in the logarithmic scale, and the solid and dash lines are used to represent repulsive and attractive forces, respectively. It is found that these two cases exhibit similar total Casimir forces in the *x*-direction shown in the blue lines. The total Casimir force is attractive at distances shorter than $\frac{d}{R} \sim 6$ due to the dominance of the equilibrium Casimir force while it is repulsive at longer distances due to the dominance of nonequilibrium Casimir force. This sign change of total Casimir force has been shown in other material systems [60].

We also show the two contributions to the total nonequilibrium Casimir force $F_x^{(2),neq} = F_{1,x}^{(2),neq}(T_{env}, T_1) + F_{2,x}^{(2),neq}(T_{env}, T_2)$ separately. These two forces $F_{1,x}^{(2),neq}(T_{env}, T_1)$ and $F_{2,x}^{(2),neq}(T_{env}, T_1)$ correspond to the total nonequilibrium Casimir force in the cases of $T_1 = 300$ K, $T_2 = 0$ K, and $T_1 = 0$ K, $T_2 = 300$ K, respectively. The repulsive nonequilibrium Casimir force on sphere 2 due to the thermal emission in sphere 1 $F_{1,x}^{(2),neq}(T_{env}, T_1)$ monotonically increases as the separation becomes smaller and below $\frac{d}{R} \sim 3.3$ it becomes attractive. Although the scaling differs in the near field, we found the scaling of $\sim d^{-2}$ at longer distances. This scaling of the force at large separations corroborates with the semi-analytical expression derived for isotropic and reciprocal spheres [60], which indicates that the off-diagonal elements of the dielectric function are of little importance in the far-field. This can be also understood from the fact that the swirling pattern of thermal emission from a sphere due to non-zero antisymmetric off diagonal elements of the dielectric function tensor fades in the far-field. The self-force $F_{2,x}^{(2),neq}(T_{env}, T_2)$ shows the oscillating behavior as a function of the separation, i.e., attractive and repulsive forces appear alternatively as a function of the inter-sphere separation. This oscillation is due to interferences of two coherent traveling waves: a wave emitted from the sphere 2 that is reflected back and passes the sphere 2 and a wave emitted from the sphere 2 in the opposite direction [60]. Constructive or destructive interferences of these waves result in different signs of the forces.

The lateral nonequilibrium Casimir forces on sphere 2 for the cases (a) and (d) are shown in Fig. 4 (c) and (f), respectively. As we expect from the qualitative discussion above, the lateral nonequilibrium Casimir forces on sphere 1 and 2 have the opposite and same signs in the cases (c) and (f), respectively. For the case of the parallel $b$ vectors, the lateral nonequilibrium Casimir force shows increasing trend as the separation becomes smaller except around $\frac{d}{R} \sim 11$ where the force $F_{1,y}^{(2),neq}$ due to thermal emission from sphere 1 and the self-force $F_{2,y}^{(2),neq}$, which have the opposite signs, become comparable. However, the force $F_{1,y}^{(2),neq}$ is always larger than $F_{2,y}^{(2),neq}$, thus the lateral nonequilibrium Casimir force acts in the same direction in the separation range we considered and in the model parameters we selected. In the case (d), however, the self-force becomes larger at around $\frac{d}{R} \lesssim 7$ and $\frac{d}{R} \gtrsim 12$ and the lateral nonequilibrium Casimir forces are attractive at short and long distances but repulsive at $7 \lesssim \frac{d}{R} \lesssim 12$. It is found in both cases that the lateral nonequilibrium Casimir forces are about one order of magnitude smaller than the equilibrium Casimir force in the *x*-direction at around $\frac{d}{R}$~7, it may be possible to realize the lateral nonequilibrium Casimir force of the same order of magnitude by optimizing the sphere sizes and material properties. In both cases, the force does not exist in the *z*-direction. Also, the two spheres in which we calculate the forces are fixed in space and are at the steady state. The detailed study of the kinematics of the spheres requires calculation of the Casimir forces between two spheres in relative motion [81].

As mentioned in Section II.D, when magnetic Weyl semimetals are ferromagnets, spontaneous magnetizations can create magnetic forces and dominate over Casimir forces. In Fig. 4 (g), we consider total Casimir forces between a magnetic Weyl semimetal sphere and a silica sphere as a system where the magnetic forces are negligible since silica do not possess permanent magnetic dipole moments. In this case, it is expected that the Casimir forces are dominant over the magnetic forces even when the magnetic Weyl semimetal is ferromagnetic. Moreover, the lateral nonequilibrium Casimir forces are expected to exist due to a swirling thermal emission from the magnetic Weyl semimetal sphere. Figure 4 (h) and (i) show the *x* and *y*-component total Casimir forces on the Weyl semimetal sphere $F_{x,y}^{(weyl)}$ and on the silica sphere $F_{x,y}^{(silica)}$. In Fig. 4 (h), it is found that the equilibrium Casimir forces on both spheres are the same magnitude and attractive, i.e., $F_x^{(weyl),eq} = -F_x^{(silica),eq} > 0$, but the nonequilibrium Casimir forces are different $F_{x,y}^{(weyl),neq} \neq F_{x,y}^{(silica),neq}$ due to the lack of symmetry that relates them even when the temperatures of the two spheres are the same at 300 K. The total force acting on the Weyl semimetal sphere in the *x*-direction shows a similar trend to the case of the two magnetic Weyl semimetal spheres, i.e., attractive force and repulsive forces at shorter and longer separations. The total Casimir force on the silica sphere, however, becomes attractive at longer distances around $\frac{d}{R}$~13-14 due to the dominance of the self-force over the force due to radiation from the Weyl semimetal sphere in certain separations. For comparison, the *x*-component total Casimir force between two silica spheres $F_x^{(silica-silica)}$ is also shown in Fig. 4 (h). It is found that the equilibrium Casimir force between two silica spheres is smaller due to smaller $\varepsilon(i\xi)$, but the nonequilibrium Casimir force is larger than that between the Weyl semimetal and silica spheres. The lateral nonequilibrium Casimir force on the Weyl semimetal sphere increases monotonically as the separation becomes smaller while the lateral force on the silica sphere becomes attractive and repulsive alternatively as a function of the separation due to the dominance of the self-force.

The lateral nonequilibrium Casimir force does not always exist even in nonreciprocal and thermal nonequilibrium cases. Figure 5 shows the configuration where the $b$ vectors in both spheres are parallel and align with the inter-sphere direction. The system is rotationally symmetric about the $x$-axis, thus the force is non-zero only in the $x$-direction. Moreover, the system is mirror-symmetric about the $yz$-plane, therefore the $z$-component force on the spheres 1 and 2 are related as $F_x^{(1)} = -F_x^{(2)}$. The numerical results in Fig. 5 show that the total Casimir force in the $z$-direction is attractive at distances shorter than $\frac{d}{R} \sim 7$ while it is repulsive at longer distances, which is a similar trend to the force observed in Fig. 4. Although not shown, the configuration where the $b$ vectors point along the $z$-axis but are antiparallel also shows the similar Casimir force in the $x$-direction and the notable difference from the result in Fig. 5 is due to the difference in the self-force $F_{2,x}^{(2),neq}$. In addition to the attractive and repulsive forces described above, the transfer of angular momentum of thermal radiation, i.e., thermal Casimir torques, can occur in the configurations in Fig. 5, as thermal radiation carries non-zero angular momentum in the direction of gyration which was recently discussed in the system of two InSb nanocubes under external magnetic fields [67].

We found above that the equilibrium Casimir force in the four configurations always acts along the inter-sphere direction and no force acts in the transverse directions. However, we expect that the way electromagnetic waves are scattered between the spheres will be different depending on the configurations of the $b$ vectors, and the Casimir energy will be configuration-dependent even when the inter-sphere separation is the same. Therefore, orientations of the $b$ vectors are expected to change to minimize the Casimir energy by exchanging momentum between the spheres.

To reveal energetically favorable system configurations at thermal equilibrium at 0 K, we study the Casimir energy dependence on the directions of the $b$ vectors. Figure 6 (a) shows a schematic of two magnetic Weyl semimetal spheres when the $b$ vectors are parallel and antiparallel and are tilted by the angle $\theta$ from the $x$-axis. Figure 6 (c) shows the corresponding Casimir energy. Note that the energy is negative since it is the energy difference from the reference state where the two spheres are infinitely far away. For both cases of the parallel and antiparallel $b$ vectors, the two spheres are most stable when the $b$ vectors are perpendicular to the inter-sphere direction.

Moreover, the parallel configuration is more stable than the antiparallel configuration. Contrarily, the system is unstable when the *b* vectors point to the inter-sphere direction.

When the two orientations of the *b* vectors are not in the energy minimum state, the Casimir force, as well as torque are exchanged between the spheres so that the orientations of the *b* vectors become parallel and perpendicular to the inter-sphere direction. To show this, in Fig. 6 (b), we consider the same configuration as the case of the parallel *b* vectors in Fig. 6 (a). We calculate the Casimir force on the outer sphere in Fig. 6 (b) for different polar angles $\alpha$ while keeping the inter-sphere distance constant at 3μm. The correspondence between the angle $\theta$ in Fig. 6 (a) and the polar angle $\alpha$ is also shown in Fig. 6 (b). The Casimir force in the direction of the polar angle $F_\alpha$ acting on the outer sphere is shown in Fig. 6 (d). As shown, the points $\alpha = 0$ or $\pi$ are mechanically stable locations. When the outer sphere deviates from these points, the Casimir force $F_\alpha$ acts as restoring force to push the outer sphere back to these points. Although the Casimir force $F_\alpha$ vanishes when the outer sphere is located exactly at $\alpha = \frac{\pi}{2}$ or $\frac{3\pi}{2}$, the configuration is mechanically unstable. Thus, a slight deviation of the outer sphere from these points results in the Casimir force $F_\alpha$ to push away the sphere until the sphere reaches the points $\alpha = 0$ or $\pi$. When the *b* vectors are perpendicular to the inter-sphere direction but are antiparallel, the Casimir torque acts between the spheres so that the *b* vectors become parallel. This Casimir torque transfer has been recently discussed in the case of two InSb nanocubes [67]. Note that this transfer of momentum in thermal equilibrium occurs only between the objects without transferring net angular momentum to the surrounding, thereby no thermodynamic law is violated.

### C. Three-sphere systems

When the number of magnetic Weyl semimetal spheres in the system is greater than two, the number of possible scattering events increases and the scattered waves interfere with each other. As a result, such a system is expected to exhibit complex dynamics. To demonstrate the applicability of the developed formalisms to many-body systems, among many possible many-body configurations of magnetic Wey semimetal spheres, we study the system consisting of three magnetic Weyl semimetal spheres situated at the vertices of an equilateral triangle. We consider the case where the *b* vectors of the three spheres are parallel and point normal to the plane of the triangle as schematically shown in Fig. 7 (a). Figure 7 (b) shows the *x*- and *y*-components of total

Casimir force on sphere 1. We found that the $z$-component force is zero both in thermal equilibrium and nonequilibrium. When the system is at global thermal equilibrium at 0K, the attractive equilibrium Casimir force on the sphere 1 $F_y^{(1),eq}$ acts in the positive $y$-direction. Due to the rotational symmetry about the $z$-axis, the equilibrium Casimir forces on spheres 2 and 3 also act towards the center of the equilateral triangle, i.e., the center of mass of the system. In thermal nonequilibrium, where the temperatures of the three spheres are at $T_1 = T_2 = T_3 = 300K$ while the environment is 0K, the nonequilibrium Casimir forces contribute to the total force. For the $y$-component force on sphere 1, the total Casimir force is attractive at separations shorter than $\frac{d}{R} \sim 6$ due to the dominance of equilibrium Casimir force and is repulsive at longer separations due to the nonequilibrium Casimir force. In addition, the lateral nonequilibrium Casimir force in the $x$-direction acts on sphere 1. From the rotational symmetry of the system, the same lateral nonequilibrium Casimir force acts on spheres 2 and 3 so that the three spheres experience the thrust to rotate globally about the center of mass of the system. In thermal nonequilibrium, thermal radiation from the three spheres carries net angular momentum to the environment. This rotational force on the three spheres is considered as a recoil force as a result of the angular momentum conservation.

Lastly, we discuss the Casimir energy of the three-sphere system in thermal equilibrium at 0 K. We fix the inter-sphere distance to $d = 3\mu m$ and change the orientations of the **b** vectors. Figure 8 (a-c) show the systems where the **b** vectors in the three spheres are in the plane of the equilateral triangle, i.e., $xy$-plane, and tilted by the angle $\theta$ measured from the $x$-axis. Case (a) is when the **b** vectors are all parallel and the Casimir energy shows little variation, which indicates that the orientations can vary under a small perturbation on the **b** vectors. Case (b) is when one of the three **b** vectors is antiparallel. In this case, the Casimir energy takes minimum at the angle around $\theta = 150°$. When we consider the case where two of the three **b** vectors are antiparallel (case (c)), the Casimir energy takes the minimum at the angle around $\theta = 90°$. Overall, we found that the cases (a-c) show a smaller variation of the Casimir energy compared to the cases where the **b** vectors are normal to the plane of the triangle as shown in the cases (d-f). Similar to the two-sphere results, the spheres are most stable when all of the three **b** vectors are parallel. The cases (e) and (f) takes the same Casimir energy at the angle $\theta = 90°$ because the Casimir energy is invariant under the time-reversal operation, i.e., flipping all **b** vectors. Note that these energy minima may not be the global minimum. In general, one needs to study the Casimir energy by changing each of the three *b* vectors independently.

## IV.  Conclusions

In summary, we derived the compact formulas for the thermal Casimir force for systems comprising an arbitrary number of spheres with arbitrary dielectric function tensors. The previous formalisms give thermal Casimir forces for arbitrary-shaped bodies but are restricted only in reciprocal systems. The derived formulas of thermal Casimir force, combined with the scattering formalism for Casimir force, allow us to study total Casimir forces among many spheres both in thermal equilibrium and nonequilibrium without the restriction of Lorentz reciprocity. Using the developed formalism, total Casimir forces between two magnetic Weyl semimetal spheres were studied for different orientations of the momentum-space Weyl node separation vectors *b* in the spheres. We also studied total Casimir forces between a magnetic Weyl semimetal sphere and a silica sphere as a case where magnetic forces due to spontaneous magnetization are negligible. In thermal nonequilibrium, we found that in addition to the total Casimir forces in the inter-sphere direction, net lateral thermal Casimir force can exist when at least one of the spheres is nonreciprocal and the *b* vector is misaligned from the inter-sphere direction. This lateral thermal Casimir force is due to angular-momentum-carrying thermal radiation from nonreciprocal spheres. We found that depending on the directions of the *b* vectors in the magnetic Weyl semimetal spheres, the spheres can experience thrust that leads to translational, rotational, and spinning motions in thermal nonequilibrium. We also discussed the dependence of the Casimir energy of the two-sphere system on the orientations of the *b* vectors and found that the *b* vectors being normal to the inter-sphere direction are more stable than other orientations. We showed that the dependence of the Casimir energy on the orientations of the *b* vectors results in the lateral Casimir force even in thermal equilibrium between the spheres to relax the system to the energy minimum state. While the transfer of Casimir torque in thermal equilibrium was predicted before, we showed that lateral Casimir force also acts in thermal equilibrium. To demonstrate the applicability of the developed formalisms for many-body nonreciprocal systems, we studied the Casimir force interactions between the three magnetic Weyl semimetal spheres situated at the vertices of an equilateral triangle. Especially, we considered the configuration of the *b* vectors where the system has the rotational symmetry. We showed that the spheres in thermal nonequilibrium experience global rotation around the center of mass of the system due to lateral thermal Casimir force. We also studied the Casimir energy in the three-sphere system and found that the ***b*** vectors being normal to the inter-sphere direction are energetically favorable, similar to the two-sphere system.

As illustrated by our study of Casimir forces in two and three magnetic Weyl semimetal spheres, this work provides a framework to study Casimir forces in many-body systems composed of both reciprocal and nonreciprocal spheres in both thermal equilibrium and nonequilibrium. With the derived formulas, one can study material properties and geometries of spheres to maximize lateral Casimir forces compared to the Casimir forces acting in the inter-sphere direction. Especially, measurements of the lateral force in thermal nonequilibrium may provide a useful method for observations of thermal Casimir force [18].

## Acknowledgments

We are grateful to Professor Shanhui Fan, Mr. Cheng Guo, and Mr. Haiwen Wang at Stanford University and Professor Linxiao Zhu at Pennsylvania State University for fruitful discussions. This work is supported by ARO MURI (Grant No. W911NF-19-1-0279) via U. Michigan. Computations of Casimir forces using MATLAB are partly performed using the Stampede2 supercomputer operated by the Extreme Science and Engineering Discovery Environment (XSEDE) through allocation No. TG-PHY200027.

* Corresponding author: ytsuri@mit.edu

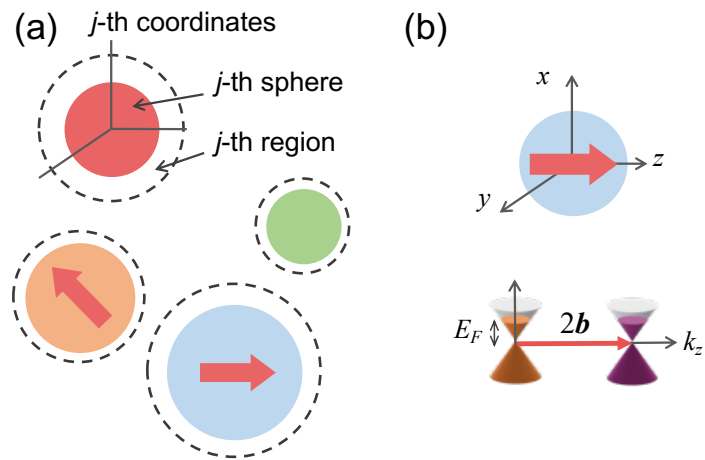

**Figure 1**: (a): *N*-sphere system in vacuum where each sphere is in local thermal equilibrium at temperature $T_j$ ($j = 1, \cdots, N$), exchanging heat and momentum radiatively between the spheres and the environment at $T_{env}$. The *j*-th coordinates have the origin at the center of the *j*-th sphere. The *j*-th region is defined as the inside of an imaginary sphere (dashed line) that encloses only the *j*-th sphere. The arrows overlapped on the spheres describe the gyration axes when spheres are gyroelectric. (b): a magnetic Weyl semimetal sphere with two Weyl nodes of opposite chirality. The Weyl nodes are separated by $2|\mathbf{b}|$ in the $k_z$-direction in the momentum space (bottom figure), which corresponds to the gyration in the *z*-axis in the real space (top figure). The Fermi energy is $E_F$ measured from the Weyl nodes.

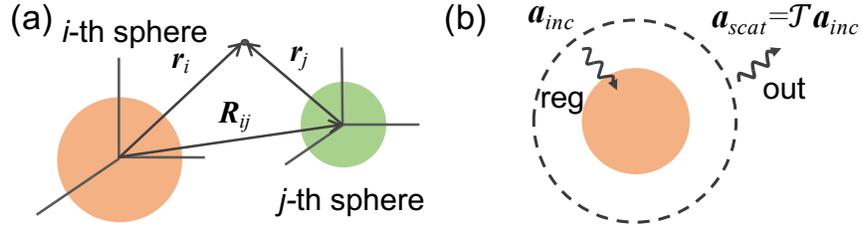

**Figure 2**: (a) A translation matrix $\mathcal{U}^{ij}(\mathbf{R}_{ij})$ describes an expansion of a wave in terms of the *j*-th coordinates (right sphere) to the expansion of the wave in terms of the *i*-th coordinates (left sphere). (b): A transition matrix relates an incident regular wave weighed by a vector of expansion coefficients $\boldsymbol{a}_{inc}$ to a scattered outgoing wave weighted a vector of expansion coefficients $\boldsymbol{a}_{scat}$.

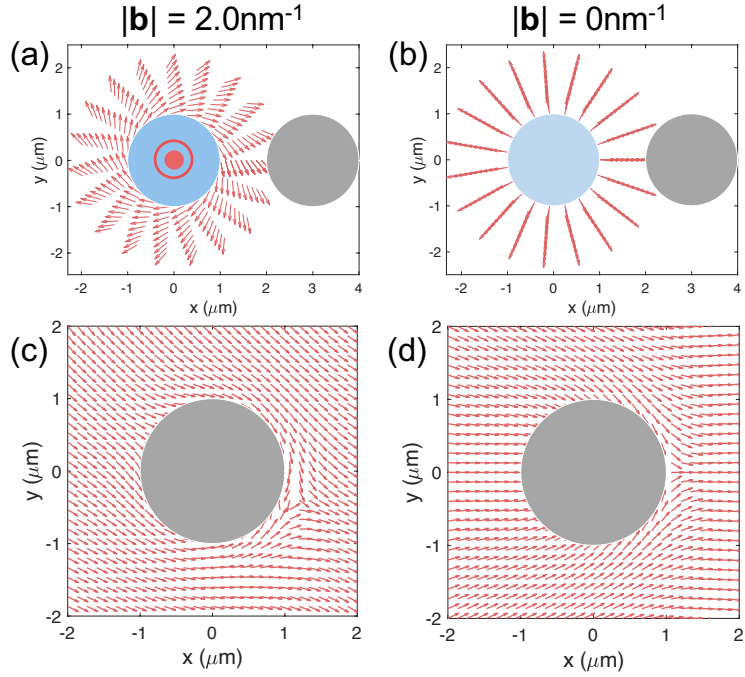

**Figure 3**: Spatial Poynting vector distribution in the system comprising of a magnetic Weyl semimetal sphere (blue) with the gyration directions in the *z*-axis as indicated by the sign on the sphere and a second sphere, which can be either reciprocal or nonreciprocal (gray). The radius of the spheres is 1μm, the inter-sphere separation is 3μm, and the temperatures are at 300 K. (a, b): Poynting vector distributions due to thermal emission from the magnetic Weyl semimetal sphere (blue) when $|b| = 2.0\text{nm}^{-1}$ (a) and $|b| = 0\text{nm}^{-1}$ (b). (c, d): Poynting vector distribution around the gray sphere due to thermal emission from the magnetic Weyl semimetal sphere when $|b| = 2.0\text{nm}^{-1}$ (c) and $|b| = 0\text{nm}^{-1}$ (d). The frequency-integrated Poynting vectors in the *z*=0 plane are plotted for all the panels, and the magnitude is normalized, thereby only their directions are meaningful.

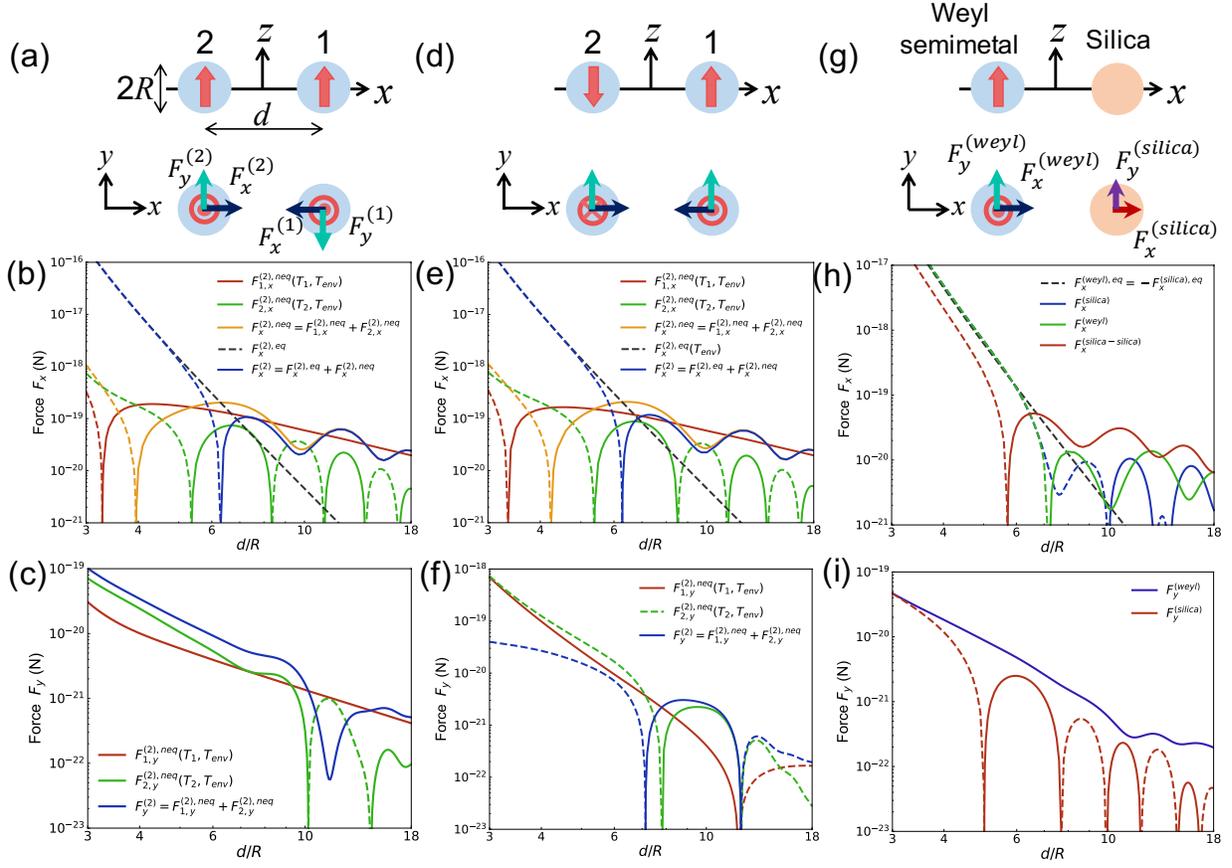

**Figure 4**: Casimir forces between two magnetic Weyl semimetal spheres with parallel (a-c) and antiparallel (d-f) gyration directions perpendicular to the inter-sphere direction, and between a magnetic Weyl semimetal sphere and a silica sphere (g-i). (a, d, g): schematics of the systems (top) and directions of the Casimir forces (bottom). In (g), different colors of the arrows indicate the forces can be different in both magnitude and direction. (b, c, e, f, h, i): The $x$- and $y$-component Casimir forces on the sphere 2 as a function of the ratio of the inter-sphere separation $d$ to the radius of the spheres $R = 1\mu m$. The panels (b,c), (e,f) and (h,i) correspond to the configurations (a), (d), and (g), respectively. The absolute value of the force is shown. For the $x$-component force, the dashed and solid lines correspond to attractive (force acts in the positive $x$-direction) and repulsive (force acts in the negative $x$-direction) forces, respectively. For the $y$-component force, the solid and dashed lines correspond to positive (force acts in the positive $y$-direction) and negative (force acts in the negative $y$-direction) forces. For the equilibrium Casimir forces, $T_{env} = 0$ K. For the nonequilibrium Casimir forces, the spheres are at $T_1 = T_2 = 300$ K while the
37

environment is at $T_{env} = 0$ K. The total Casimir forces are the summations of the equilibrium and nonequilibrium Casimir forces as defined in Eq. (1).

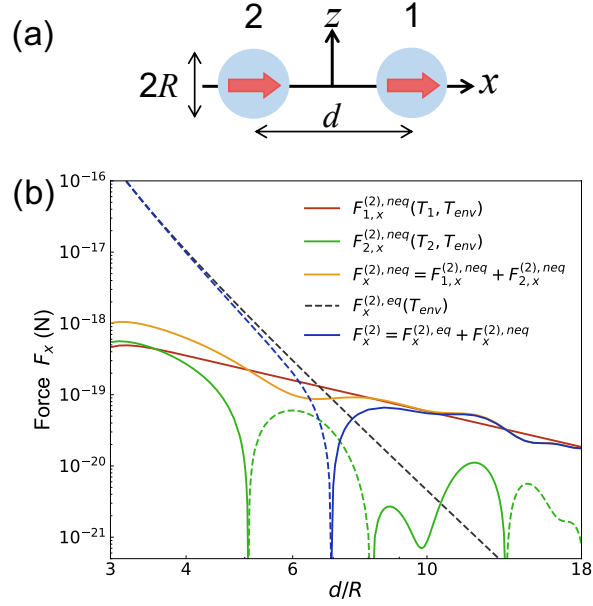

**Figure 5**: (a): A schematic of two magnetic Weyl semimetal spheres when the gyration axes of the spheres are parallel and aligned with the inter-sphere direction. (b): Casimir forces in the $z$-direction on the sphere 2 as a function of the ratio of the inter-sphere distance $d$ to the radii of the spheres $R = 1\mu m$. The force acting on the sphere 1 is obtained by the symmetry $F_x^{(1)} = -F_x^{(2)}$. The absolute values of the forces are shown in the logarithmic scale. The dashed and solid lines correspond to attractive (forces on the sphere 2 are in the positive $x$-direction) and repulsive (forces on the sphere 2 are in the negative $x$-direction) forces, respectively. The $y$- and $z$-component forces are zero. The temperatures of the spheres and the environment are at $T_1 = T_2 = 300$ K and $T_{env} = 0$ K, respectively.

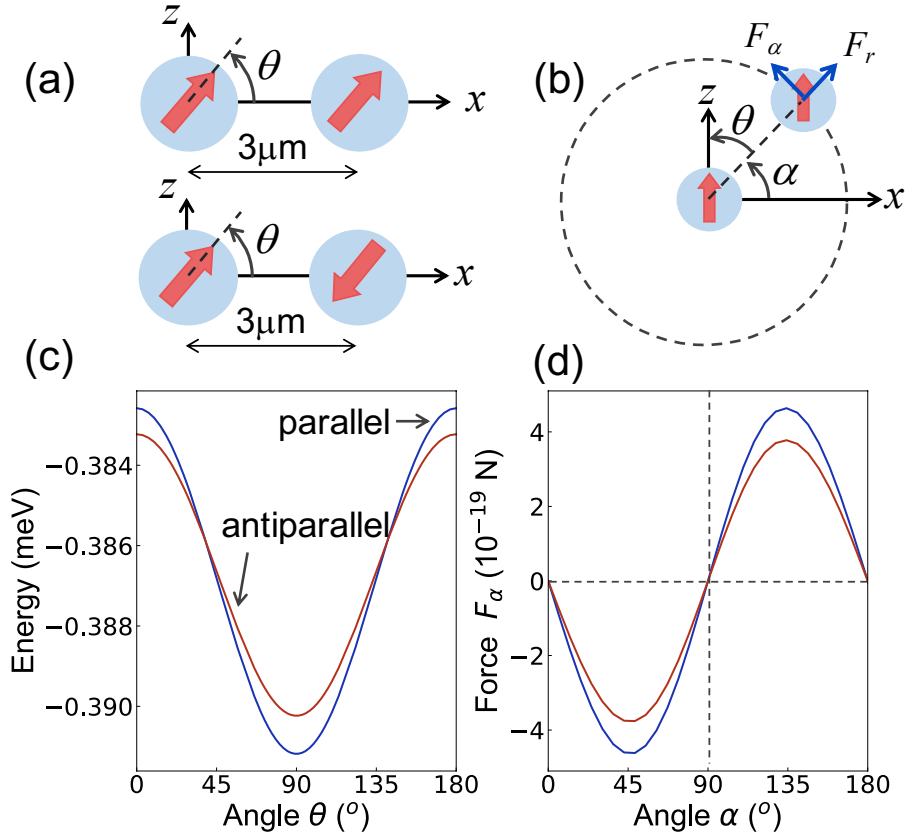

**Figure 6**: Casimir energy and force dependence on the angle of the gyration axes in two magnetic Weyl semimetal spheres at global thermal equilibrium at $T_1 = T_2 = T_{env} = 0$ K. (a): A schematic of two parallel and antiparallel gyration axes tilted by the angle $\theta$ measured from the *x*-axis. The angle $\theta$ varies in *xz*-plane. (b): An equivalent schematic to the parallel gyration axes in (a). The angle $\theta$ in (b) is equivalent to the angle $\theta$ in (a). (c): Casimir energy dependence on the angle $\theta$ of the parallel (blue line) and antiparallel (red line) gyration axes in two magnetic Weyl semimetal spheres. (d): Casimir force in the polar angle direction $F_\alpha$ acting on the outer sphere when the outer sphere moves along the circle (dashed line). The two angles are related by $\alpha = \frac{\pi}{2} - \theta$. The gyration axes of the two sphere are fixed in the *z*-axis. In (c) and (d), the radii of the spheres are 1μm and the inter-sphere separation is fixed to 3μm.

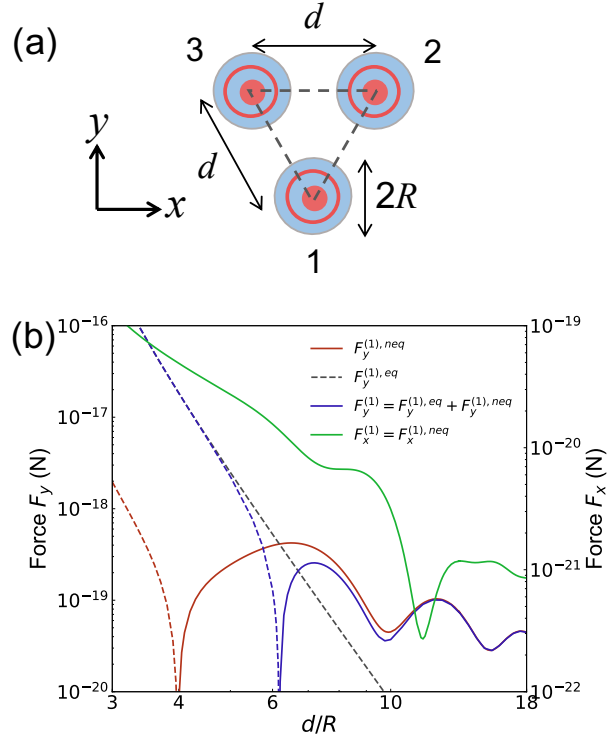

**Figure 7**: (a): A schematic of three magnetic Weyl semimetal spheres when the gyration axes of the three spheres are parallel and normal to the plane of the equilateral triangle (dashed line) pointing to the positive *z*-direction. (b): Casimir forces acting on the sphere 1 as a function of the ratio of the inter-sphere separation *d* to the radii of the spheres $R = 1\mu m$. The Casimir forces on the spheres 2 and 3 can be inferred from the rotational symmetry about the *z*-axis of the system. The absolute values of the forces are plotted in the logarithmic scale. The dashed and solid lines correspond to positive (the *x*- and *y*-component forces point in the positive *x* and *y* directions, respectively) and negative (the *x*- and *y*-component forces point in the negative *x* and *y* directions, respectively) forces, respectively. The *z*-component force is zero. The temperatures of the spheres and the environment are $T_1 = T_2 = T_3 = 300K$ and $T_{env} = 0$ K, respectively.

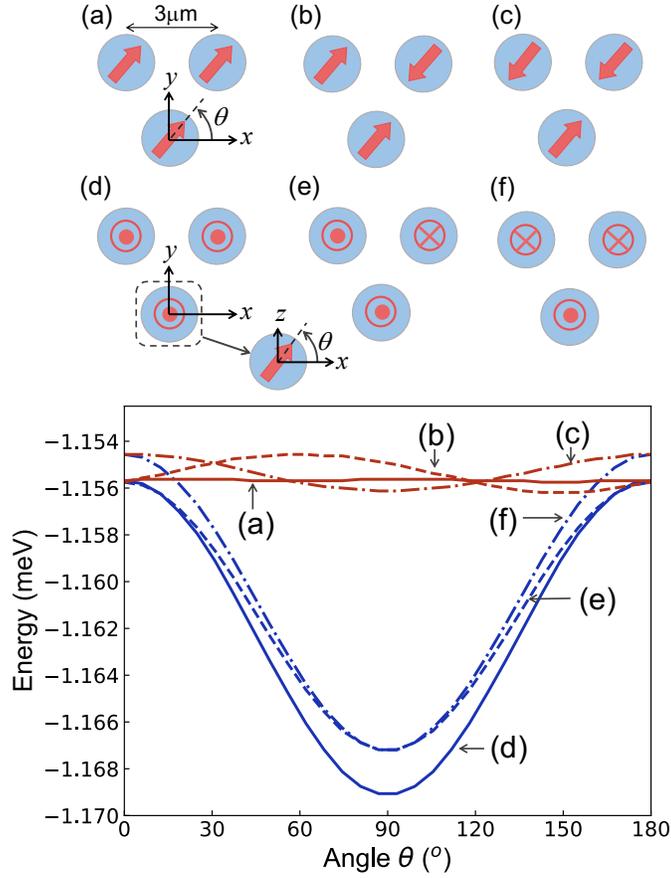

**Figure 8**: Casimir energy dependence on the angle $\theta$ of the gyration axes in three magnetic Weyl semimetal spheres situated at the vertices of an equilateral triangle. The radii of the spheres are 1μm and the inter-sphere separation fixed to 3μm. (a-c): The gyration axes are in the *xy*-plane and the angle $\theta$ is measured from the *x*-axis. (a-c) correspond to the cases where all of the three **b** vectors are parallel (a), one of them is antiparallel (b), and two of them are antiparallel (c). (d-f): The gyration axes are normal to the *xy*-plane and the angle $\theta$ is measured from the *x*-axis in *xz*-plane. (d-f) correspond to the cases where all three vectors **b** are parallel (d) and one of them is antiparallel (e), and two of them are antiparallel (f).

# Appendices

## Appendix A: Equilibrium Casimir force

### 1. Vector spherical wave basis

The vector spherical wave basis for TE polarization defined in the imaginary frequency is [55]:

$$\boldsymbol{E}^{\alpha}_{(M,l,m)}(\kappa,\boldsymbol{r}) = \nabla \times \frac{z_l^{(\alpha)}(\kappa r) Y_{lm}(\theta,\phi)\boldsymbol{r}}{\sqrt{l(l+1)}}, \tag{A1}$$

where $\kappa = -ik_0$ is the imaginary wavenumber. The radial function $z_l^{\alpha}(\kappa r)$ is the modified spherical Bessel function of the first kind when $\alpha = reg$, i.e., $z_l^{(reg)}(z) = i_l(z) = \sqrt{\frac{\pi}{2z}} I_{l+\frac{1}{2}}(z)$, and the modified spherical Bessel function of the third kind when $\alpha = out$, i.e., $z_l^{(out)}(z) = k_l(z) = \sqrt{\frac{2}{\pi z}} K_{l+\frac{1}{2}}(z)$. The vector spherical wave basis for TM polarization is defined as:

$$\boldsymbol{E}^{\alpha}_{(N,l,m)}(\kappa,\boldsymbol{r}) = \frac{1}{\kappa} \nabla \times \boldsymbol{E}^{\alpha}_{(M,l,m)}(\kappa,\boldsymbol{r}), \tag{A2}$$

In the spherical coordinates, the free dyadic Green's function is given for $|\boldsymbol{r}| > |\boldsymbol{r}'|$ by:

$$G_0(\boldsymbol{r},\boldsymbol{r}',ic\kappa) = \sum_{P,l,m} \left[ C_P(\kappa) \boldsymbol{E}^{out}_{(P,l,m)}(\kappa,\boldsymbol{r}) \otimes \boldsymbol{E}^{reg*}_{(P,l,m)}(\kappa,\boldsymbol{r}') \right], \tag{A3}$$

where the coefficients for the two polarizations are $C_M(\kappa) = -C_N(\kappa) = \kappa$.

### 2. Translation matrix

The matrix elements of the translation matrix between the $i$-th and $j$-th objects that are outside of one other, $\mathcal{U}^{ji}$ are given by [55]:

$$\mathcal{U}^{ji}_{(M,l',m'),(M,l,m)} = (-1)^{m+l} \sum_p [l(l+1) + l'(l'+1) - p(p+1)] \times$$

$$\sqrt{\frac{\pi(2l+1)(2l'+1)(2p+1)}{l(l+1)l'(l'+1)}} \times$$

$$\begin{pmatrix} l & l' & p \\ 0 & 0 & 0 \end{pmatrix} \begin{pmatrix} l & l' & p \\ m & -m' & m'-m \end{pmatrix} k_p(\kappa|\boldsymbol{R}_{ji}|) Y_{p,m-m'}(\hat{\boldsymbol{R}}_{ji}), \tag{A4}$$

$$\mathcal{U}^{ji}_{(N,l',m')(M,l,m)} = -\frac{i\kappa}{\sqrt{l(l+1)l'(l'+1)}} \mathbf{r}_{ji} \cdot \begin{bmatrix} \mathbf{e}_x \frac{1}{2} \begin{pmatrix} \lambda^+_{l,m} A_{(l',m'),(l,m+1)}(\mathbf{R}_{ji}) + \\ \lambda^-_{l,m} A_{(l',m'),(l,m-1)}(\mathbf{R}_{ji}) \end{pmatrix} \\ +\mathbf{e}_y \frac{1}{2i} \begin{pmatrix} \lambda^+_{l,m} A_{(l',m'),(l,m+1)}(\mathbf{R}_{ji}) - \\ \lambda^-_{l,m} A_{(l',m'),(l,m-1)}(\mathbf{R}_{ji}) \end{pmatrix} \\ +\mathbf{e}_z m A_{(l',m'),(l,m)}(\mathbf{R}_{ji}) \end{bmatrix}, \quad (A5)$$

$$\mathcal{U}^{ji}_{(M,l',m')(N,l,m)} = -\mathcal{U}^{ji}_{(N,l',m')(M,l,m)}, \quad (A6)$$

$$\mathcal{U}^{ji}_{(N,l',m')(N,l,m)} = \mathcal{U}^{ji}_{(M,l',m')(M,l,m)}, \quad (A7)$$

where

$$A_{(l',m'),(l,m)}(\mathbf{r}_{ji}) = (-1)^{m+l} \sum_p \sqrt{4\pi(2l+1)(2l'+1)(2p+1)} \times$$
$$\begin{pmatrix} l & l' & p \\ 0 & 0 & 0 \end{pmatrix} \begin{pmatrix} l & l' & p \\ m & -m' & m'-m \end{pmatrix} k_p(\kappa|\mathbf{R}_{ji}|) Y_{p,m-m'}(\widehat{\mathbf{R}}_{ji}), \quad (A8)$$

and $\lambda^\pm_{l,m} = \sqrt{(l \mp m)(l \pm m + 1)}$. $\mathbf{e}_x$, $\mathbf{e}_y$, $\mathbf{e}_z$ are the unit vectors in the Cartesian coordinates. In Eq, (A8), the brackets are Wigner 3j symbols.

### 3. T-matrix in the imaginary frequency

The T-matrix of a magnetic Weyl semimetal sphere in the real frequency is obtained by expanding the incident, scattered, and internal waves about the sphere via vector spherical waves and then by matching the coefficients in the wave expansions so the boundary conditions are satisfied [61,77]. The T-matrix obtained in this way is used for computations of nonequilibrium Casimir force in our work. The T-matrix in the imaginary frequency domain can be obtained by rewriting the real frequency with the imaginary frequency. In doing so, we use the relations between the spherical Bessel functions and modified spherical Bessel functions: $i_l(x) = i^{-l} j_l(ix)$ and $k_l(x) = -i^l h_l^{(1)}(ix)$ where $i$ is the unit imaginary number. Also, in calculating the T-matrix, it is convenient to define the functions $D_l^{(1)}(z) = \frac{[zi_l(z)]'}{zi_l(z)}$ and $D_l^{(3)}(z) = \frac{[zk_l(z)]'}{zk_l(z)}$. The function $D_l^{(1)}(z)$ is calculated by the downward recurrence relation

$$D_{l-1}^{(1)}(z) = \frac{1}{D_l^{(1)}(z) + \frac{l}{z}} + \frac{l}{z}, \quad (A9)$$

using the initial value $D^{(1)}_{n_{st}}(z) = 0$ with $n_{st} = \max(|z|, l_{max}) + 15$ where $l_{max}$ is the cutoff of the polynomial degree. The function $D_l^{(3)}(z)$ is calculated by upward recurrent relation:

$$D_l^{(3)}(z) = \frac{1}{D_{l-1}^{(3)}(z) - \frac{l}{z}} - \frac{l}{z}, \tag{A10}$$

starting with $D_0^{(1)}(z) = -1$.

## Appendix B: Nonequilibrium Casimir force

### 1. Vector spherical wave basis

The vector spherical waves in this work follow those defined in [59] and are given as:

$$\boldsymbol{E}^\alpha_{(M,l,m)}(\boldsymbol{r}, \omega) = \sqrt{k_0(-1)^m} z_l^{(\alpha)}(k_0 r) \boldsymbol{V}^{(2)}_{l,m}(r, \theta, \phi), \tag{B1}$$

$$\boldsymbol{E}^\alpha_{(N,l,m)}(\boldsymbol{r}, \omega) = \sqrt{k_0(-1)^m} \left[ \sqrt{l(l+1)} \frac{z_l^{(\alpha)}(k_0 r)}{kr} \boldsymbol{V}^{(1)}_{l,m}(r, \theta, \phi) + \zeta_l^{(\alpha)}(k_0 r) \boldsymbol{V}^{(3)}_{l,m}(r, \theta, \phi) \right], \tag{B2}$$

where $r, \theta, \phi$ are the radial distance, polar angle, and azimuthal angle in the spherical coordinates, respectively, and $k_0 = \omega/c$ is the magnitude of the wavevector in vacuum. $\alpha$ is the label for regular ($\alpha$=reg) and outgoing ($\alpha$=out) waves with which the radial function $z_l^{(\alpha)}(kr)$ is the spherical Bessel function of the first kind $j_l(x) = \sqrt{\frac{\pi}{2x}} J_{l+\frac{1}{2}}(x)$ and the spherical Hankel function of the first kind $h_l^{(1)}(x) = \sqrt{\frac{\pi}{2x}} H^{(1)}_{l+\frac{1}{2}}(x)$, respectively. We introduced the function $\zeta_l^{(\alpha)}(x) = \frac{1}{x} \frac{d}{dx}\left[ x z_l^{(\alpha)}(x) \right]$. The polarizations $N$ and $M$ represent TM and TE modes, respectively, and $l$ and $m$ are the labels of the spherical harmonics. These two polarizations are related by

$$\begin{aligned} k_0 \boldsymbol{E}^\alpha_{(M,l,m)} &= \nabla \times \boldsymbol{E}^\alpha_{(N,l,m)}, \\ k_0 \boldsymbol{E}^\alpha_{(N,l,m)} &= \nabla \times \boldsymbol{E}^\alpha_{(M,l,m)}. \end{aligned} \tag{B3}$$

We define the spherical harmonic vector functions in Eqs. (B1) and (B2) as:

$$\boldsymbol{V}^{(1)}_{l,m}(r, \theta, \phi) = \boldsymbol{e}_r Y_{l,m}(\theta, \phi), \tag{B4}$$

$$\boldsymbol{V}^{(2)}_{l,m}(\boldsymbol{r},\theta,\phi) = \nabla \times \frac{[rY_{l,m}(\theta,\phi)]}{\sqrt{l(l+1)}} = \frac{1}{\sqrt{l(l+1)}}\left[\frac{im}{\sin\theta}Y_{l,m}\boldsymbol{e}_\theta - \frac{\partial Y_{l,m}}{\partial \theta}\boldsymbol{e}_\phi\right], \quad (B5)$$

$$\boldsymbol{V}^{(3)}_{l,m}(\boldsymbol{r},\theta,\phi) = \frac{r\nabla Y_{l,m}(\theta,\phi)}{\sqrt{l(l+1)}} = \frac{1}{\sqrt{l(l+1)}}\left[\frac{\partial Y_{l,m}}{\partial \theta}\boldsymbol{e}_\theta + \frac{im}{\sin\theta}Y_{l,m}\boldsymbol{e}_\phi\right], \quad (B6)$$

where $Y_{l,m}$ is the spherical harmonics and we use the standard definition as in [82]. $\boldsymbol{e}_r, \boldsymbol{e}_\theta, \boldsymbol{e}_\phi$ are the unit vectors in the spherical coordinates. Unlike the conventional definitions [76,82] of the spherical waves, we have a factor $\sqrt{k_0(-1)^m}$ in Eqs. (B1) and (B2). With this definition, the free Green's function $G(\omega,\boldsymbol{r},\boldsymbol{r}')$ for $|\boldsymbol{r}| > |\boldsymbol{r}'|$ is given as Eq. (16) in the main text. Also, the complex conjugate of the regular wave basis satisfies $\boldsymbol{E}^{reg*}_{(P,l,m)} = \boldsymbol{E}^{reg}_{(P,l,-m)}$ (P=N, M) when using the identity $\left(\sqrt{(-1)^m}\right)^* = (-1)^m\sqrt{(-1)^m}$.

## 2. Matrix elements in the force components

In this section, we give the explicit expressions of the matrix elements in Eqs. (34), (37), (39) in the main text. Although not shown here, the identities in [69,83] are found useful in our derivations. In the following results, we define and use the auxiliary functions $I^{(j)}$ ($j = 1,\cdots,9$), $Q_{(l,m)(l',m')}$, and $J_{(l,m)(l',m')}$. The explicit expressions are given at the end of this section.

The matrices $\mathcal{B}^{(\alpha,\beta)}$ and $\overline{\mathcal{B}}^{(\alpha,\beta)}$ are defined as

$$\mathcal{B}^{(\alpha,\beta)} \equiv \begin{bmatrix} \left[\mathcal{B}^{(\alpha,\beta)}_{(N,l,m),(N,l',m')}\right] & \left[\mathcal{B}^{(\alpha,\beta)}_{(N,l,m),(M,l',m')}\right] \\ \left[\mathcal{B}^{(\alpha,\beta)}_{(M,l,m),(N,l',m')}\right] & \left[\mathcal{B}^{(\alpha,\beta)}_{(M,l,m),(M,l',m')}\right] \end{bmatrix}, \quad (B7)$$

$$\overline{\mathcal{B}}^{(\alpha,\beta)} \equiv \begin{bmatrix} \left[\mathcal{B}^{(\alpha,\beta)}_{(M,l,m),(M,l',m')}\right] & \left[\mathcal{B}^{(\alpha,\beta)}_{(M,l,m),(N,l',m')}\right] \\ \left[\mathcal{B}^{(\alpha,\beta)}_{(N,l,m),(M,l',m')}\right] & \left[\mathcal{B}^{(\alpha,\beta)}_{(N,l,m),(N,l',m')}\right] \end{bmatrix}, \quad (B8)$$

where the matrix elements are given as:

$$\mathcal{B}^{(\alpha,\beta)}_{(N,l,m),(N,l',m')}(x) = (-1)^{m'}\sqrt{(-1)^m}\sqrt{(-1)^{m'}}\delta_{m,m'} \times$$

$$\left[\begin{array}{l} z_l^{(\alpha)}(x)z_{l'}^{(\beta)*}(x)\sqrt{l(l+1)}\sqrt{l'(l'+1)} \times \\ \left[\begin{array}{l} \delta_{l,l'+1}\sqrt{\dfrac{(l'+1)^2 - m'^2}{(2l'+1)(2l'+3)}} \\ +\delta_{l,l'-1}\sqrt{\dfrac{l'^2 - m'^2}{(2l'-1)(2l'+1)}} \end{array}\right] \\ +x^2\zeta_l^{(\alpha)}(x)\zeta_{l'}^{(\beta)*}(x) \times \\ \left[\begin{array}{l} \delta_{l,l'+1}\dfrac{\sqrt{l'(l'+2)}}{l'+1}\sqrt{\dfrac{(l'+1)^2 - m'^2}{(2l'+1)(2l'+3)}} \\ +\delta_{l,l'-1}\dfrac{\sqrt{l'^2-1}}{l'}\sqrt{\dfrac{l'^2-m'^2}{(2l'-1)(2l'+1)}} \end{array}\right] \end{array}\right], \quad (B9)$$

$$\mathcal{B}^{(\alpha,\beta)}_{(N,l,m),(M,l',m')}(x) = (-1)^{m'}\sqrt{(-1)^m}\sqrt{(-1)^{m'}} \times$$
$$x^2\zeta_l^{(\alpha)}(x)z_{l'}^{(\beta)*}(x)\frac{-im}{l(l+1)}\delta_{l,l'}\delta_{m,m'} \quad (B10)$$

$$\mathcal{B}^{(\alpha,\beta)}_{(M,l,m),(N,l',m')}(x) = (-1)^{m'}\sqrt{(-1)^m}\sqrt{(-1)^{m'}} \times$$
$$x^2 z_l^{(\alpha)}(x)\zeta_{l'}^{(\beta)*}(x)\frac{im}{l(l+1)}\delta_{l,l'}\delta_{m,m'}, \quad (B11)$$

$$\mathcal{B}^{(\alpha,\beta)}_{(M,l,m),(M,l',m')}(x) = x^2 z_l^\alpha(x)z_{l'}^{\beta*}(x) \times$$

$$\left[\begin{array}{l} \delta_{l,l'+1}\dfrac{\sqrt{l'(l'+2)}}{l'+1}\sqrt{\dfrac{(l'+1)^2 - m'^2}{(2l'+1)(2l'+3)}} \\ +\delta_{l,l'-1}\dfrac{\sqrt{l'^2-1}}{l'}\sqrt{\dfrac{l'^2-m'^2}{(2l'-1)(2l'+1)}} \end{array}\right]\delta_{m,m'}. \quad (B12)$$

The matrix $\mathcal{C}^{(\alpha,\beta)}$ is defined as

$$\mathcal{C}^{(\alpha,\beta)} \equiv \begin{bmatrix} \left[\mathcal{C}^{(\alpha,\beta)}_{(N,l,m),(N,l',m')}\right] & \left[\mathcal{C}^{(\alpha,\beta)}_{(M,l,m),(N,l',m')}\right] \\ \left[\mathcal{C}^{(\alpha,\beta)}_{(M,l,m),(N,l',m')}\right] & \left[\mathcal{C}^{(\alpha,\beta)}_{(N,l,m),(N,l',m')}\right] \end{bmatrix}, \quad (B13)$$

where the matrix elements are given as:

$$\mathcal{C}^{(\alpha,\beta)}_{(M,l,m),(N,l',m')}(x) = imx z_l^{(\alpha)}(x) z_{l'}^{(\beta)*}(x) \delta_{l,l'} \delta_{m,m'} \tag{B14}$$

$$\mathcal{C}^{(\alpha,\beta)}_{(N,l,m),(N,l',m')}(x) = x \zeta_l^{\alpha}(x) z_{l'}^{(\beta)*}(x) \times$$

$$\begin{bmatrix} -\delta_{l,l'+1}(l'+2)\sqrt{\dfrac{l'}{l'+2}}\sqrt{\dfrac{(l'+1)^2 - m^2}{(2l'+1)(2l'+3)}} \\ +\delta_{l,l'-1}(l'-1)\sqrt{\dfrac{l'+1}{l'-1}}\sqrt{\dfrac{l'^2 - m^2}{(2l'-1)(2l'+1)}} \end{bmatrix} \delta_{m,m'}. \tag{B15}$$

The matrix $\mathcal{D}^{(\alpha,\beta)}$ is defined as:

$$\mathcal{D}^{(\alpha,\beta)} \equiv \begin{bmatrix} \left[\mathcal{D}^{(\alpha,\beta)}_{(N,l,m),(N,l',m')}\right] & [0] \\ [0] & \left[\mathcal{D}^{(\alpha,\beta)}_{(N,l,m),(N,l',m')}\right] \end{bmatrix}, \tag{B16}$$

where the matrix elements are given as:

$$\mathcal{D}^{(\alpha,\beta)}_{(N,l,m)(N,l',m')}(x) = (-1)^{m'}\sqrt{(-1)^m}\sqrt{(-1)^{m'}} \times$$
$$\sqrt{l(l+1)l'(l'+1)}\, z_l^{(\alpha)}(x) z_{l'}^{(\beta)*}(x) I^{(7)}_{(l,m)(l',m')}. \tag{B17}$$

The matrices $\mathcal{E}^{(\alpha,\beta)}$ and $\bar{\mathcal{E}}^{(\alpha,\beta)}$ are defined as:

$$\mathcal{E}^{(\alpha,\beta)} \equiv \begin{bmatrix} \left[\mathcal{E}^{(\alpha,\beta)}_{(N,l,m),(N,l',m')}\right] & \left[\mathcal{E}^{(\alpha,\beta)}_{(N,l,m),(M,l',m')}\right] \\ \left[\mathcal{E}^{(\alpha,\beta)}_{(M,l,m),(N,l',m')}\right] & \left[\mathcal{E}^{(\alpha,\beta)}_{(M,l,m),(M,l',m')}\right] \end{bmatrix}, \tag{B18}$$

$$\bar{\mathcal{E}}^{(\alpha,\beta)} \equiv \begin{bmatrix} \left[\mathcal{E}^{(\alpha,\beta)}_{(M,l,m),(M,l',m')}\right] & \left[\mathcal{E}^{(\alpha,\beta)}_{(M,l,m),(N,l',m')}\right] \\ \left[\mathcal{E}^{(\alpha,\beta)}_{(N,l,m),(M,l',m')}\right] & \left[\mathcal{E}^{(\alpha,\beta)}_{(N,l,m),(N,l',m')}\right] \end{bmatrix}, \tag{B19}$$

where the matrix elements are given as:

$$\mathcal{E}^{(\alpha,\beta)}_{(N,l,m),(N,l',m')}(x) = (-1)^{m'}\sqrt{(-1)^m}\sqrt{(-1)^{m'}} \times$$
$$\begin{bmatrix} \sqrt{l(l+1)l'(l'+1)}\, z_l^{(\alpha)}(x) z_{l'}^{(\beta)*}(x) I^{(7)}_{(l,m),(l',m')} \\ + \dfrac{x^2 \zeta_l^{(\alpha)}(x) \zeta_{l'}^{(\beta)*}(x)}{\sqrt{l(l+1)l'(l'+1)}} I^{(8)}_{(l,m),(l',m')} \end{bmatrix}, \tag{B20}$$

$$\mathcal{E}^{(\alpha,\beta)}_{(N,l,m),(M,l',m')}(x) = -i(-1)^{m'}\sqrt{(-1)^m}\sqrt{(-1)^{m'}}\frac{x^2\zeta_l^{(\alpha)}(x)z_{l'}^{(\beta)*}(x)}{\sqrt{l(l+1)l'(l'+1)}} \times$$

$$\left[ m\sqrt{(l'-m')(l'+m'+1)}\left( \begin{array}{l} -\sqrt{\dfrac{(l+m+1)(l+m+2)}{(2l+1)(2l+3)}}I^{(6)}_{(l+1,m+1)(l',m'+1)} \\ +\sqrt{\dfrac{(l-m)(l-m-1)}{(2l+1)(2l-1)}}I^{(6)}_{(l-1,m+1)(l',m'+1)} \end{array} \right) \right.$$

$$+2mm'\left( \begin{array}{l} \sqrt{\dfrac{(l+m+1)(l-m+1)}{(2l+1)(2l+3)}}I^{(6)}_{(l+1,m)(l',m')} \\ +\sqrt{\dfrac{(l+m)(l-m)}{(2l+1)(2l-1)}}I^{(6)}_{(l-1,m)(l',m')} \end{array} \right)$$

$$\left. +m'\sqrt{(l-m)(l+m+1)}\left( \begin{array}{l} -\sqrt{\dfrac{(l'+m'+1)(l'+m'+2)}{(2l'+1)(2l'+3)}}I^{(6)}_{(l'+1,m'+1)(l,m+1)} \\ +\sqrt{\dfrac{(l'-m')(l'-m'-1)}{(2l'+1)(2l'-1)}}I^{(6)}_{(l'-1,m'+1)(l,m+1)} \end{array} \right) \right], \quad \text{(B21)}$$

$$\mathcal{E}^{(\alpha,\beta)}_{(M,l,m),(N,l',m')}(x) = i(-1)^{m'}\sqrt{(-1)^m}\sqrt{(-1)^{m'}}\frac{x^2z_l^{(\alpha)}(x)\zeta_{l'}^{(\beta)*}(x)}{\sqrt{l(l+1)l'(l'+1)}} \times$$

$$\left[ m\sqrt{(l'-m')(l'+m'+1)}\left( \begin{array}{l} -\sqrt{\dfrac{(l+m+1)(l+m+2)}{(2l+1)(2l+3)}}I^{(6)}_{(l+1,m+1)(l',m'+1)} \\ +\sqrt{\dfrac{(l-m)(l-m-1)}{(2l+1)(2l-1)}}I^{(6)}_{(l-1,m+1)(l',m'+1)} \end{array} \right) \right.$$

$$+2mm'\left( \begin{array}{l} \sqrt{\dfrac{(l+m+1)(l-m+1)}{(2l+1)(2l+3)}}I^{(6)}_{(l+1,m)(l',m')} \\ +\sqrt{\dfrac{(l+m)(l-m)}{(2l+1)(2l-1)}}I^{(6)}_{(l-1,m)(l',m')} \end{array} \right)$$

$$\left. +m'\sqrt{(l-m)(l+m+1)}\left( \begin{array}{l} -\sqrt{\dfrac{(l'+m'+1)(l'+m'+2)}{(2l'+1)(2l'+3)}}I^{(6)}_{(l'+1,m'+1)(l,m+1)} \\ +\sqrt{\dfrac{(l'-m')(l'-m'-1)}{(2l'+1)(2l'-1)}}I^{(6)}_{(l'-1,m'+1)(l,m+1)} \end{array} \right) \right], \quad \text{(B22)}$$

$$\mathcal{E}^{(\alpha,\beta)}_{(M,l,m),(M,l',m')}(x) = (-1)^{m'}\sqrt{(-1)^m}\sqrt{(-1)^{m'}} \times$$
$$\frac{x^2 z_l^{(\alpha)}(x) z_{l'}^{(\beta)*}(x)}{\sqrt{l(l+1)l'(l'+1)}} I^{(8)}_{(l,m)(l',m')}. \tag{B23}$$

The matrix $\mathcal{F}^{(\alpha,\beta)}$ is defined as:

$$\mathcal{F}^{(\alpha,\beta)} \equiv \begin{bmatrix} \left[\mathcal{F}^{(\alpha,\beta)}_{(N,l,m),(N,l',m')}\right] & \left[\mathcal{F}^{(\alpha,\beta)}_{(M,l,m),(N,l',m')}\right] \\ \left[\mathcal{F}^{(\alpha,\beta)}_{(M,l,m),(N,l',m')}\right] & \left[\mathcal{F}^{(\alpha,\beta)}_{(N,l,m),(N,l',m')}\right] \end{bmatrix}, \tag{B24}$$

where the matrix elements are given as:

$$\mathcal{F}^{(\alpha,\beta)}_{(N,l,m),(N,l',m')}(x) = (-1)^{m'}\sqrt{(-1)^m}\sqrt{(-1)^{m'}} \times$$
$$x\zeta_l^{(\alpha)}(x) z_{l'}^{(\beta)*}(x)\sqrt{\frac{l'(l'+1)}{l(l+1)}} \times$$
$$\left[\sqrt{(l-m)(l+m+1)}I^{(2)}_{(l,m)(l',m')} + m I^{(1)}_{(l,m)(l',m')}\right], \tag{B25}$$

$$\mathcal{F}^{(\alpha,\beta)}_{(M,l,m),(N,l',m')}(x) = (-1)^{m'}\sqrt{(-1)^m}\sqrt{(-1)^{m'}} x z_l^{(\alpha)}(x) z_{l'}^{(\beta)*}(x) im\sqrt{\frac{l'(l'+1)}{l(l+1)}} \times$$
$$\left[\sqrt{\frac{(l+m+1)(l-m+1)}{(2l+1)(2l+3)}}I^{(6)}_{(l+1,m)(l',m')} + \sqrt{\frac{(l+m)(l-m)}{(2l+1)(2l-1)}}I^{(6)}_{(l-1,m)(l',m')}\right]. \tag{B26}$$

The matrix $\mathcal{G}^{(\alpha,\beta)}$ is defined as:

$$\mathcal{G}^{(\alpha,\beta)} \equiv \begin{bmatrix} \left[\mathcal{G}^{(\alpha,\beta)}_{(N,l,m),(N,l',m')}\right] & \left[\mathcal{G}^{(\alpha,\beta)}_{(M,l,m),(N,l',m')}\right] \\ \left[\mathcal{G}^{(\alpha,\beta)}_{(M,l,m),(N,l',m')}\right] & \left[\mathcal{G}^{(\alpha,\beta)}_{(N,l,m),(N,l',m')}\right] \end{bmatrix}, \tag{B27}$$

where the matrix elements are given as:

$$\mathcal{G}^{(\alpha,\beta)}_{(N,l,m)(N,l',m')}(x) = (-1)^{m'}\sqrt{(-1)^m}\sqrt{(-1)^{m'}} \times$$
$$x\zeta_l^{(\alpha)} z_{l'}^{(\beta)*}(x) im \sqrt{\frac{l'(l'+1)}{l(l+1)}} I^{(3)}_{(l,m)(l',m')}, \tag{B28}$$

$$\mathcal{G}^{(\alpha,\beta)}_{(M,l,m)(N,l',m')}(x) = -(-1)^{m'}\sqrt{(-1)^m}\sqrt{(-1)^{m'}} \times$$

$$xz_l^{(\alpha)}(x)z_{l'}^{(\beta)*}(x)\sqrt{\frac{l'(l'+1)}{l(l+1)}} \times$$

$$\left[\sqrt{(l-m)(l+m+1)}\left\{\begin{array}{c}\sqrt{\frac{(l-m)(l-m+1)}{(2l+1)(2l+3)}}I^{(3)}_{(l+1,m)(l',m')} \\ -\sqrt{\frac{(l+m)(l+m+1)}{(2l+1)(2l-1)}}I^{(3)}_{(l-1,m)(l',m')}\end{array}\right\}\right.$$

$$\left.+m\left\{\begin{array}{c}\sqrt{\frac{(l+m+1)(l-m+1)}{(2l+1)(2l+3)}}I^{(3)}_{(l+1,m)(l',m')} \\ +\sqrt{\frac{(l+m)(l-m)}{(2l+1)(2l-1)}}I^{(3)}_{(l-1,m)(l',m')}\end{array}\right\}\right]. \quad (B29)$$

The matrix $\mathcal{H}^{(\alpha,\beta)}$ is defined as:

$$\mathcal{H}^{(\alpha,\beta)} \equiv \begin{bmatrix}\left[\mathcal{H}^{(\alpha,\beta)}_{(N,l,m),(N,l',m')}\right] & [0] \\ [0] & \left[\mathcal{H}^{(\alpha,\beta)}_{(N,l,m),(N,l',m')}\right]\end{bmatrix}, \quad (B30)$$

where the matrix elements are given as:

$$\mathcal{H}^{(\alpha,\beta)}_{(N,l,m),(N,l',m')}(x) = \sqrt{(-1)^m}\sqrt{(-1)^{m'}}(-1)^{m'}\sqrt{l(l+1)l'(l'+1)} \times$$
$$z_l^{(\alpha)}(x)z_{l'}^{(\beta)*}(x)\left[I^{(3)}_{(l,m)(l',m')} - I^{(4)}_{(l,m)(l',m')}\right]. \quad (B31)$$

The matrices $\mathcal{J}^{(\alpha,\beta)}$ and $\bar{\mathcal{J}}^{(\alpha,\beta)}$ is defined as:

$$\mathcal{J}^{(\alpha,\beta)} = \begin{bmatrix}\left[\mathcal{J}^{(\alpha,\beta)}_{(N,l,m),(N,l',m')}\right] & \left[\mathcal{J}^{(\alpha,\beta)}_{(N,l,m),(M,l',m')}\right] \\ \left[\mathcal{J}^{(\alpha,\beta)}_{(M,l,m),(N,l',m')}\right] & \left[\mathcal{J}^{(\alpha,\beta)}_{(M,l,m),(M,l',m')}\right]\end{bmatrix}, \quad (B32)$$

$$\bar{\mathcal{J}}^{(\alpha,\beta)} = \begin{bmatrix}\left[\mathcal{J}^{(\alpha,\beta)}_{(M,l,m),(M,l',m')}\right] & \left[\mathcal{J}^{(\alpha,\beta)}_{(M,l,m),(N,l',m')}\right] \\ \left[\mathcal{J}^{(\alpha,\beta)}_{(N,l,m),(M,l',m')}\right] & \left[\mathcal{J}^{(\alpha,\beta)}_{(N,l,m),(N,l',m')}\right]\end{bmatrix}, \quad (B33)$$

where the matrix elements are given as:

$$\mathcal{J}^{(\alpha,\beta)}_{(N,l,m)(N,l',m')}(x) = \sqrt{(-1)^m}\sqrt{(-1)^{m'}}(-1)^{m'}\begin{bmatrix} \sqrt{l(l+1)l'(l'+1)}z_l^{(\alpha)}(x)z_{l'}^{(\beta)*}(x) \times \\ \left(I^{(3)}_{(l,m)(l',m')} - I^{(4)}_{(l,m)(l',m')}\right) \\ + \dfrac{x^2 \zeta_l^{(\alpha)}(x)\zeta_{l'}^{(\beta)*}(x)}{\sqrt{l(l+1)l'(l'+1)}} I^{(9)}_{(l,m)(l',m')} \end{bmatrix}, \quad (B34)$$

$$\mathcal{J}^{(\alpha,\beta)}_{(N,l,m)(M,l',m')}(x) = -\sqrt{(-1)^m}\sqrt{(-1)^{m'}}(-1)^{m'} \frac{x^2 \zeta_l^{(\alpha)}(x)z_{l'}^{(\beta)*}(x)}{\sqrt{l(l+1)l'(l'+1)}} \times$$

$$\begin{bmatrix} im\sqrt{(l'-m')(l'+m'+1)}\left(\begin{array}{l} -\sqrt{\dfrac{(l+m+1)(l+m+2)}{(2l+1)(2l+3)}} I^{(3)}_{(l+1,m+1)(l',m'+1)} \\ +\sqrt{\dfrac{(l-m)(l-m-1)}{(2l-1)(2l+1)}} I^{(3)}_{(l-1,m+1)(l',m'+1)} \end{array}\right) \\ +2imm'\left(\begin{array}{l} \sqrt{\dfrac{(l+m+1)(l-m+1)}{(2l+1)(2l+3)}} I^{(3)}_{(l+1,m)(l',m')} \\ +\sqrt{\dfrac{(l+m)(l-m)}{(2l-1)(2l+1)}} I^{(3)}_{(l-1,m)(l',m')} \end{array}\right) \\ im'\sqrt{(l-m)(l+m+1)}\left(\begin{array}{l} -\sqrt{\dfrac{(l'+m'+1)(l'+m'+2)}{(2l'+1)(2l'+3)}} I^{(3)}_{(l,m+1)(l'+1,m'+1)} \\ +\sqrt{\dfrac{(l'-m')(l'-m'-1)}{(2l'-1)(2l'+1)}} I^{(3)}_{(l,m+1)(l'-1,m'+1)} \end{array}\right) \end{bmatrix}, \quad (B35)$$

$$\mathcal{J}^{(\alpha,\beta)}_{(M,l,m)(N,l',m')}(x) = \sqrt{(-1)^m}\sqrt{(-1)^{m'}}(-1)^{m'} \frac{x^2 z_l^{(\alpha)}(x)\zeta_{l'}^{(\beta)*}(x)}{\sqrt{l(l+1)l'(l'+1)}} \times$$

$$\begin{bmatrix} im\sqrt{(l'-m')(l'+m'+1)}\left(\begin{array}{l}-\sqrt{\frac{(l+m+1)(l+m+2)}{(2l+1)(2l+3)}}I^{(3)}_{(l+1,m+1)(l',m'+1)} \\ +\sqrt{\frac{(l-m)(l-m-1)}{(2l-1)(2l+1)}}I^{(3)}_{(l-1,m+1)(l',m'+1)}\end{array}\right) \\ +2imm'\left(\begin{array}{l}\sqrt{\frac{(l+m+1)(l-m+1)}{(2l+1)(2l+3)}}I^{(3)}_{(l+1,m)(l',m')} \\ +\sqrt{\frac{(l+m)(l-m)}{(2l-1)(2l+1)}}I^{(3)}_{(l-1,m)(l',m')}\end{array}\right) \\ im'\sqrt{(l-m)(l+m+1)}\left(\begin{array}{l}-\sqrt{\frac{(l'+m'+1)(l'+m'+2)}{(2l'+1)(2l'+3)}}I^{(3)}_{(l,m+1)(l'+1,m'+1)} \\ +\sqrt{\frac{(l'-m')(l'-m'-1)}{(2l'-1)(2l'+1)}}I^{(3)}_{(l,m+1)(l'-1,m'+1)}\end{array}\right) \end{bmatrix}, \quad (B36)$$

$$\mathcal{J}^{(\alpha,\beta)}_{(M,l,m)(M,l',m')}(x) = \sqrt{(-1)^m}\sqrt{(-1)^{m'}}(-1)^{m'} \frac{x^2 z_l^{(\alpha)}(x)z_{l'}^{(\beta)*}(x)}{\sqrt{l(l+1)l'(l'+1)}} \times$$

$$\begin{bmatrix} \sqrt{(l-m)(l+m+1)(l'-m')(l'+m'+1)}\left(I^{(3)}_{(l,m+1)(l',m'+1)} - I^{(4)}_{(l,m+1)(l',m'+1)}\right) \\ +m'\sqrt{(l-m)(l+m+1)}I^{(5)}_{(l,m)(l',m')} + m\sqrt{(l'-m')(l'+m'+1)}I^{(5)*}_{(l',m')(l,m)} \\ +mm'\left(I^{(3)}_{(l,m)(l',m')} + I^{(4)}_{(l,m)(l',m')}\right) \end{bmatrix}. \quad (B37)$$

The matrix $\mathcal{K}^{(\alpha,\beta)}$ is defined as:

$$\mathcal{K}^{(\alpha,\beta)} \equiv \begin{bmatrix} \left[\mathcal{K}^{(\alpha,\beta)}_{(N,l,m),(N,l',m')}\right] & \left[\mathcal{K}^{(\alpha,\beta)}_{(M,l,m),(N,l',m')}\right] \\ \left[\mathcal{K}^{(\alpha,\beta)}_{(M,l,m),(N,l',m')}\right] & \left[\mathcal{K}^{(\alpha,\beta)}_{(N,l,m),(N,l',m')}\right] \end{bmatrix}, \quad (B38)$$

where the matrix elements are given as:

$$\mathcal{K}^{(\alpha,\beta)}_{(N,l,m)(N,l',m')}(x) = \sqrt{(-1)^m}\sqrt{(-1)^{m'}}(-1)^{m'} x\zeta_l^{(\alpha)}(x)z_{l'}^{(\beta)*}(x)\sqrt{\frac{l'(l'+1)}{l(l+1)}} \times$$

$$\left[\sqrt{(l-m)(l+m+1)}I^{(5)}_{(l,m)(l',m')} + mI^{(4)}_{(l,m)(l',m')}\right], \quad (B39)$$

$$\mathcal{K}^{(\alpha,\beta)}_{(M,l,m)(N,l',m')}(x) = i\sqrt{(-1)^m}\sqrt{(-1)^{m'}}(-1)^{m'} \times$$

$$mxz_l^{(\alpha)}(x)z_{l'}^{(\beta)*}(x)\sqrt{\frac{l'(l'+1)}{l(l+1)}} \times$$

$$\left[\begin{array}{l} \sqrt{\dfrac{(l+m+1)(l-m+1)}{(2l+1)(2l+3)}} I^{(3)}_{(l+1,m)(l',m')} \\ + \sqrt{\dfrac{(l+m)(l-m)}{(2l-1)(2l+1)}} I^{(3)}_{(l-1,m)(l',m')} \end{array}\right]. \tag{B40}$$

The matrix $\mathcal{L}^{(\alpha,\beta)}$ is defined as:

$$\mathcal{L}^{(\alpha,\beta)} \equiv \begin{bmatrix} \left[\mathcal{L}^{(\alpha,\beta)}_{(N,l,m),(N,l',m')}\right] & \left[\mathcal{L}^{(\alpha,\beta)}_{(M,l,m),(N,l',m')}\right] \\ \left[\mathcal{L}^{(\alpha,\beta)}_{(M,l,m),(N,l',m')}\right] & \left[\mathcal{L}^{(\alpha,\beta)}_{(N,l,m),(N,l',m')}\right] \end{bmatrix}, \tag{B41}$$

where the matrix elements are given as:

$$\mathcal{L}^{(\alpha,\beta)}_{(N,l,m)(N,l',m')}(x) = i\sqrt{(-1)^m}\sqrt{(-1)^{m'}}(-1)^{m'} \times$$

$$mx\zeta_l^{(\alpha)}(x)z_{l'}^{(\beta)*}(x)\sqrt{\frac{l'(l'+1)}{l(l+1)}} I^{(6)}_{(l,m)(l',m')}, \tag{B42}$$

$$\mathcal{L}^{(\alpha,\beta)}_{(M,l,m)(N,l',m')}(x) = -\sqrt{(-1)^m}\sqrt{(-1)^{m'}}(-1)^{m'} xz_l^{(\alpha)}(x)z_{l'}^{(\beta)*}(x)\sqrt{\frac{l'(l'+1)}{l(l+1)}} \times$$

$$\left[\begin{array}{l} \sqrt{(l-m)(l+m+1)}\left(\begin{array}{l} \sqrt{\dfrac{(l-m)(l-m+1)}{(2l+1)(2l+3)}} I^{(6)}_{(l+1,m)(l',m')} \\ - \sqrt{\dfrac{(l+m)(l+m+1)}{(2l+1)(2l+3)}} I^{(6)}_{(l-1,m)(l',m')} \end{array}\right) \\ +m\left(\begin{array}{l} \sqrt{\dfrac{(l+m+1)(l-m+1)}{(2l+1)(2l+3)}} I^{(6)}_{(l+1,m)(l',m')} \\ + \sqrt{\dfrac{(l+m)(l-m)}{(2l+1)(2l-1)}} I^{(6)}_{(l-1,m)(l',m')} \end{array}\right) \end{array}\right]. \tag{B43}$$

In the above expressions, we define the auxiliary integrals $I^{(j)}$ ($j = 1, \cdots, 9$). We found the derivation of the analytical expressions of the auxiliary functions $I^{(j)}$ ($j = 1, \cdots, 6$) in [84] useful. Below, the shorthand notation for the integral $\int d\theta d\phi \equiv \int_0^\pi d\theta \int_0^{2\pi} d\phi$ is used.

$$I^{(1)}_{(l,m)(l',m')} \equiv \int d\theta d\phi \cos^2\theta \cos\phi\, Y_{l,m} Y^*_{l',m'}$$

$$= \sqrt{\frac{(l+m)(l-m)}{(2l+1)(2l-1)}} \sqrt{\frac{(l+m-1)(l-m-1)}{(2l-3)(2l-1)}} I^{(1)'}_{(l-2,m)(l',m')}$$

$$+ \left[\frac{(l+m)(l-m)}{(2l+1)(2l-1)} + \frac{(l+m+1)(l-m+1)}{(2l+1)(2l+3)}\right] I^{(1)'}_{(l,m)(l',m')}$$

$$+ \sqrt{\frac{(l+m+1)(l-m+1)}{(2l+1)(2l+3)}} \sqrt{\frac{(l+m+2)(l-m+2)}{(2l+3)(2l+5)}} I^{(1)'}_{(l+2,m)(l',m')}, \tag{B44}$$

where $I^{(1)'}_{(l,m)(l',m')}$ is defined as

$$I^{(1)'}_{(l,m)(l',m')} = \pi \gamma_{l,m} \gamma_{l',m'} \left(\delta_{m,m'-1} + \delta_{m,m'+1}\right) Q_{(l,m)(l',m')}, \tag{B45}$$

where the auxiliary function $Q_{(l,m)(l',m')}$ is given in Eq. (B48).

$$I^{(2)}_{(lm)(l'm')} \equiv \int d\theta d\phi \cos\theta \sin\theta \cos\phi\, e^{-i\phi} Y_{l,m+1} Y^*_{l',m'}$$

$$= \pi \gamma_{l-1,m+1} \gamma_{l',m'} \sqrt{\frac{(l+m+1)(l-m-1)}{(2l+1)(2l-1)}} \delta_{m,m'-1} J_{(l-1,|m+1|)(l',m')}$$

$$+ \pi \gamma_{l-1,m+1} \gamma_{l',m'} \sqrt{\frac{(l+m+1)(l-m-1)}{(2l+1)(2l-1)}} \delta_{m,m'+1} J_{(l-1,|m+1|)(l',m')}$$

$$+ \pi \gamma_{l+1,m+1} \gamma_{l',m'} \sqrt{\frac{(l+m+2)(l-m)}{(2l+1)(2l+3)}} \delta_{m,m'-1} J_{(l+1,|m+1|)(l',m')}$$

$$+ \pi \gamma_{l+1,m+1} \gamma_{l',m'} \sqrt{\frac{(l+m+2)(l-m)}{(2l+1)(2l+3)}} \delta_{m,m'+1} J_{(l+1,|m+1|)(l',m')}, \tag{B46}$$

where the auxiliary function $J_{(l,m)(l',m')}$ is given in Eq. (57).

$$I^{(3)}_{(lm)(l'm')} \equiv \int d\theta d\phi \sin\phi\, Y_{l,m} Y^*_{l',m'}$$

$$= i\pi \gamma_{l,m} \gamma_{l',m'} \left(\delta_{m,m'+1} - \delta_{m,m'-1}\right) Q_{(l,m)(l',m')}, \tag{B47}$$

$$I^{(4)}_{(l,m)(l',m')} \equiv \int d\theta d\phi \cos^2\theta \sin\phi \, Y_{l,m} Y^*_{l',m'}$$

$$= \sqrt{\frac{(l+m)(l-m)}{(2l+1)(2l-1)}} \sqrt{\frac{(l+m-1)(l-m-1)}{(2l-3)(2l-1)}} I^{(4)'}_{(l-2,m)(l',m')}$$

$$+ \left[\frac{(l+m)(l-m)}{(2l+1)(2l-1)} + \frac{(l+m+1)(l-m+1)}{(2l+1)(2l+3)}\right] I^{(4)'}_{(l,m)(l',m')}$$

$$+ \sqrt{\frac{(l+m+1)(l-m+1)}{(2l+1)(2l+3)}} \sqrt{\frac{(l+m+2)(l-m+2)}{(2l+3)(2l+5)}} I^{(4)'}_{(l+2,m)(l',m')}, \tag{B48}$$

where

$$I^{(4)'}_{(l,m)(l',m')} = i\pi\gamma_{l,m}\gamma_{l',m'}(\delta_{m,m'+1} - \delta_{m,m'-1}) Q_{(l,m)(l',m')}. \tag{B49}$$

$$I^{(5)}_{(l,m)(l',m')} \equiv \int d\theta d\phi \cos\theta \sin\theta \sin\phi \, e^{-i\phi} Y_{l,m+1} Y^*_{l',m'}$$

$$= -i\pi\gamma_{l-1,m+1}\gamma_{l',m'} \sqrt{\frac{(l+m+1)(l-m-1)}{(2l+1)(2l-1)}} \delta_{m,m'-1} J_{(l-1,|m+1|)(l',m')}$$

$$+ i\pi\gamma_{l-1,m+1}\gamma_{l',m'} \sqrt{\frac{(l+m+1)(l-m-1)}{(2l+1)(2l-1)}} \delta_{m,m'+1} J_{(l-1,|m+1|)(l',m')}$$

$$- i\pi\gamma_{l+1,m+1}\gamma_{l',m'} \sqrt{\frac{(l+m+2)(l-m)}{(2l+1)(2l+3)}} \delta_{m,m'-1} J_{(l+1,|m+1|)(l',m')}$$

$$+ i\pi\gamma_{l+1,m+1}\gamma_{l',m'} \sqrt{\frac{(l+m+2)(l-m)}{(2l+1)(2l+3)}} \delta_{m,m'+1} J_{(l+1,|m+1|)(l',m')}. \tag{B50}$$

$$I^{(6)}_{(l,m)(l',m')} \equiv \int d\theta d\phi \cos\phi \, Y_{l,m} Y^*_{l',m'}$$

$$= \pi\gamma_{l,m}\gamma_{l',m'}(\delta_{m,m'+1} + \delta_{m,m'-1}) Q_{(l,m)(l',m')}. \tag{B51}$$

$$I^{(7)}_{(l,m)(l',m')} \equiv I^{(6)}_{(l,m)(l',m')} - I^{(1)}_{(l,m)(l',m')}. \tag{B52}$$

$$I^{(8)}_{(l,m)(l',m')} = \int d\theta d\phi \sin^2\theta \cos\phi \left(\frac{\partial Y_{lm}}{\partial \theta}\frac{\partial Y^*_{l'm'}}{\partial \theta} + \frac{mm'}{\sin^2\theta}Y_{lm}Y_{l'm'}\right)$$
$$= \sqrt{(l-m)(l+m+1)}\sqrt{(l'-m')(l'+m'+1)}I^{(7)}_{(l,m+1)(l',m'+1)}$$
$$+ m'\sqrt{(l-m)(l+m+1)}I^{(2)}_{(l,m)(l',m')} + mm'I^{(1)}_{(l,m)(l',m')}$$
$$+ m\sqrt{(l'-m')(l'+m'+1)}I^{(2)*}_{(l',m')(l,m)} + mm'I^{(1)}_{(l,m)(l',m')} + mm'I^{(7)}_{(l,m)(l',m')}. \quad (B53)$$

$$I^{(9)}_{(l,m)(l',m')} = \int d\theta d\phi \sin\theta \sin\phi \left(\frac{\partial Y_{lm}}{\partial \theta}\frac{\partial Y^*_{l'm'}}{\partial \theta} + \frac{mm'}{\sin^2\theta}Y_{lm}Y_{l'm'}\right)$$
$$\sqrt{(l-m)(l+m+1)}\sqrt{(l'-m')(l'+m'+1)} \times$$
$$\left(I^{(3)}_{(l,m+1)(l',m'+1)} - I^{(4)}_{(l,m+1)(l',m'+1)}\right) + mm'\left(I^{(3)}_{(l,m)(l',m')} + I^{(4)}_{(l,m)(l',m')}\right)$$
$$+ m'\sqrt{(l-m)(l+m+1)}I^{(5)}_{(l,m)(l',m')} + m\sqrt{(l'-m')(l'+m'+1)}I^{(5)*}_{(l,m)(l',m')}. \quad (B54)$$

The auxiliary function $Q_{(l,m)(l',m')}$ is defined as:

$$Q_{(l,m)(l',m')} \equiv \int_{-1}^{1} dx \frac{P_l^{|m|}(x)P_{l'}^{|m'|}}{\sqrt{1-x^2}}, \quad (B55)$$

where $P_l^m$ is the associated Legendre function. For our purpose, we only need the results when $m$ and $m'$ differ by 1. In this case, it can be shown from the parity of the integral that $l+l'$ must be odd for Eq. (B55) to be non-zero. Thus, writing $l' = l + 2j + 1$ with $j \in \mathbb{Z}$, we have the following identities [84,85]:

$$\int_{-1}^{1} dx \frac{P_l^{|m|}(x)P_{l+2j+1}^{|m|+1}}{\sqrt{1-x^2}} = \begin{cases} 0 & j < 0 \\ -\frac{2(l+m)!}{(l-m)!} & j \geq 0 \end{cases},$$
$$\int_{-1}^{1} dx \frac{P_l^{|m|+1}(x)P_{l+2j+1}^{|m|}}{\sqrt{1-x^2}} = \begin{cases} -\frac{2(l+|m|+2j+1)!}{(l-|m|+2j+1)!} & j < 0 \\ 0 & j \geq 0 \end{cases}. \quad (B56)$$

The auxiliary function $J_{(l,m)(l',m')}$ is the overlap function of associated Legendre functions defined as:

$$J_{(l,m)(l',m')} = \int_{-1}^{1} dx\, P_l^m(x)P_{l'}^{m'}(x), \quad (B57)$$

and we use the closed form expression derived by Wong [86] to evaluate it. It is given as:

$$J_{(l,m)(l',m')} = \sum_{p_1=0}^{p_{1max}} \sum_{p_2=0}^{p_{2max}} a_{l,m}^{p_1} a_{l',m'}^{p_2} \times$$

$$\frac{\Gamma\left(\frac{1}{2}(l+l'-m-m'-2p_1-2p_2+1)\right)\Gamma\left(\frac{1}{2}(m+m'+2p_1+2p_2+2)\right)}{\Gamma\left(\frac{1}{2}(l+l'+3)\right)}, \quad (B58)$$

for $0 \leq m \leq l$ and $0 \leq m' \leq l'$ provided that $l+l'-m-m'$ is an even integer. When $l+l'-m-m'$ is an odd integer, Eq. (B58) is zero. $\Gamma(x)$ is the gamma function. The upper limit of the summation is $p_{max} = \lfloor (l-m)/2 \rfloor$ which is the integer part of $(l-m)/2$. The coefficients are given as:

$$a_{l,m}^p = \frac{(-1)^p (l+m)!}{2^{m+2p}(m+p)!\, p!\, (l-m-2p)!}. \quad (B59)$$